\documentclass{article}
\usepackage[a4paper,top=2cm,bottom=2cm,left=2cm,right=2cm]{geometry}
\usepackage{authblk}
  \usepackage{paralist}
  \usepackage{graphics}
  \usepackage{epsfig} 
\usepackage{graphicx} 
\usepackage{epstopdf}
 \usepackage[colorlinks=true]{hyperref}
\hypersetup{urlcolor=blue, citecolor=red}
\usepackage{amsmath} 
\usepackage{amsfonts}
\usepackage[bbgreekl]{mathbbol}
\DeclareSymbolFontAlphabet{\mathbbm}{bbold}
\DeclareSymbolFontAlphabet{\mathbb}{AMSb}
\usepackage{amssymb}
\usepackage{amsthm}
\usepackage{mathrsfs}
\usepackage{stackrel,graphicx}
\usepackage{enumitem}
\usepackage{cases}
\usepackage{booktabs}
\usepackage{braket}
\usepackage[strict]{changepage}
\usepackage{subfigure}
\usepackage[small,justification=justified]{caption}
\usepackage{titlesec} 
\usepackage{multicol}
\usepackage{hyperref}
\usepackage{chemarr}

\usepackage{amssymb,amsmath,amsthm}
\usepackage{booktabs}
\usepackage{multirow}
\usepackage[table]{xcolor}
\usepackage{subfigure}
\usepackage{chemfig}
\usepackage[version=4]{mhchem}
\usepackage[right]{lineno}

\newtheorem{oss*}{Observation}
\newtheorem*{rem*}{Remark}
\theoremstyle{definition}

\newcommand{\Ub}{{\bf U}}

\newcommand{\f}{{\bf f}}
\newcommand{\bv}{{\bf v}}

\newcommand{\x}{{\bf x}}

\newcommand{\bV}{{\bf V}}

\newcommand{\RR}{\mathbb{R}}

\providecommand{\keywords}[1]{\textbf{\textit{Keywords --}} #1}

\newenvironment{sistem}
{\left\lbrace\begin{array}{@{}l@{}}}
{\end{array}\right.}

\begin{document}

\title{A kinetic derivation of spatial distributed models for tumor-immune system interactions}
\author[1]{Martina Conte\footnote{Corresponding author: martina.conte@unipr.it\\Abbreviated title: Kinetic Description of Tumor-Immune Dynamics}}
\author[1,2]{Romina Travaglini}
\affil[1]{\centerline {\small Department of Mathematical, Physical and Computer Sciences, University of Parma} \newline \centerline{\small  Parco Area delle Scienze 53/A, 43124, Parma, Italy}}
\affil[2]{\centerline{\small INDAM -- Istituto Nazionale di Alta Matematica "Francesco Severi"} 
   \newline 
   \centerline{\small Piazzale Aldo Moro 5, 00185, Roma, Italy}}
\date{\today}                     
\setcounter{Maxaffil}{0}
\renewcommand\Affilfont{\itshape\small}

\maketitle

\begin{abstract}
We propose a mathematical kinetic framework to investigate the complex nonlinear interactions between tumor cells and the immune system, focusing on the spatial dynamics that drive tumor progression and immune responses. We develop two kinetic models: the first represents a conservative scenario where immune cells switch between active and passive states without proliferation, while the second incorporates immune cell proliferation and apoptosis. By considering specific assumptions on the microscopic processes, we formally derive macroscopic systems featuring linear diffusion, nonlinear cross-diffusion, and nonlinear self-diffusion. Using dynamical systems theory, we examine the equilibrium states and their stability, and conduct numerical simulations to validate our theoretical findings. Our analysis reveals clear correspondences between the derived macroscopic systems and the underlying kinetic model. Moreover, these findings underscore the significance of spatial interactions in shaping tumor-immune dynamics, paving the way for a more structured and targeted exploration of immune responses across different pathological contexts.
\end{abstract}
 
 
\section{Introduction}\label{sec:intro}
The spread and progression of tumor cells into healthy tissue is a complex phenomenon involving multiple mechanisms across various spatial and temporal scales. Within this context, the interactions between cancer cells and the immune system are crucial in determining the tumor evolution and the effectiveness of therapeutic strategies. In fact, tumor cells may adopt numerous tactics to evade detection and destruction by immune cells, including the secretion of immunosuppressive factors, alterations in antigen presentation, and the recruitment of regulatory immune cells \cite{kim2022evasion,vinay2015immune}. Simultaneously, the immune system employs diverse mechanisms to recognize and eliminate tumor cells, such as activating cytotoxic T-lymphocytes and natural killer cells \cite{gonzalez2018roles,finn2012immuno}. In recent years, various therapeutic strategies have been developed to sustain and enhance immune responses. These include therapies based on interleukins, which are cytokines primarily expressed and secreted by leukocytes that play essential roles in immune cell activation, differentiation, proliferation, maturation, migration, and adhesion \cite{briukhovetska2021interleukins,samadi2023cancer}. Exploring immune dynamics and the complex tumor-immune interactions is essential for gaining deeper insights into tumor evolution and immune responses, including tumor adaptation to the microenvironment and the functioning of immune surveillance.

Spatial interactions significantly shape the behavior and dynamics of biological systems, particularly in tumor growth and immune responses. The spatial arrangement of cells within the tumor microenvironment influences cellular interactions, resource allocation, and overall tumor progression \cite{van2024spatial,fu2021spatial}. Understanding these spatial dynamics is critical for unraveling the complexities of tumor biology and the immune system response to cancer. Moreover, spatial interactions have important implications for therapeutic strategies. Understanding how spatial dynamics affect tumor-immune interactions can guide the development of more effective treatments. For instance, therapies designed to enhance immune cell migration into tumors or alter the spatial distribution of tumor cells could lead to improved treatment outcomes. 

Mathematical modeling plays a pivotal role in advancing our understanding of tumor-immune system dynamics by providing a structured framework to analyze their complex interactions. It allows us to translate biological phenomena into formal equations, enabling {\it in silico} analysis of different scenarios and the exploration of hypothetical outcomes. This approach has the potential to elucidate key aspects of the underlying biological mechanisms. During the last years, many papers have dealt with the problem of devising reliable dynamical models of tumor development, making use of systems of ODEs for describing the interactions between malignant and immune cells \cite{adam2012survey,de2003mathematical,de2005validated,eftimie2011interactions,robertson2012mathematical,kuznetsov1994nonlinear,conte2018qualitative,conte2025car}. However, for a better understanding of the complex dynamics of tumor growth and immune interactions, spatially distributed models are increasingly recognized as essential. Unlike traditional lumped models that treat populations as homogeneous entities, spatially distributed models account for the inherent spatial heterogeneity of biological systems. This allows for a more nuanced exploration of cellular behaviors, interactions, and responses to therapies. Moreover, these models can effectively simulate the dynamics of cell movement and migration, such as the directed motion of immune cells toward tumor sites and the invasion of surrounding tissues by tumor cells. Diffusion operators play a crucial role in modeling cell movement and, according to the expected spatial structure, different types of operators can be utilized, including linear and nonlinear diffusion, cross-diffusion, and self-diffusion terms.  The choice for their particular shape depends on the expected spatial structure. Specifically, cross-diffusion operators are applicable when dealing with interacting populations, where the diffusion of one species is influenced by the concentration gradient of another. These operators are often employed in scenarios involving spatial segregation. In contrast, self-diffusion operators pertain to a single population, where diffusion depends on the population's concentration, leading to complex and rich behaviors \cite{lou1995diffusion}. To mathematically capture the diverse observed spatial dynamics, various classes of spatially distributed models have been developed. These models are based on both discrete (or hybrid) approaches \cite{gong2017computational, mallet2006cellular, almeida2022hybrid, macfarlane2018modelling} and continuous frameworks \cite{de2006spatial, atsou2020size, Maxime2021}, where the different diffusion operators are explicitly defined at the population level. 

Recently, the application of kinetic transport equations (KTEs) has gained attraction in the study of cell movement and interactions \cite{chalub2004kinetic, conte2022multi, hillen2006m5, chauviere2007amodeling, conte2023non, kelkel2012multiscale, loy2020kinetic}, particularly in the context of tumor evolution \cite{conte2020glioma, engwer2015glioma, buckwar2023stochastic, kolbe2021modeling, lorenz2014class, chiari2025multi}. Kinetic models characterize the dynamics of distribution functions for the cell populations, incorporating, besides time and positions, various kinetic variables, such as microscopic velocity and activity levels. These models account for binary individual interactions at a microscopic level using stochastic frameworks, leading to Chapman–Kolmogorov equations within the theory of Markov processes \cite{jager1992distribution}. This approach parallels the derivation of the nonlinear Boltzmann equation found in gas kinetic theory \cite{cercignani1988boltzmann}. Boltzmann-type equations for cell population densities are analyzed, and scaling arguments are employed to derive equations for macroscopic observables, typically focusing on the first few moments of the distribution functions. The adoption of a kinetic approach to address issues in immunology is motivated not only by the deeper insights it provides, but also by its relevance to the early growth phase of tumors, characterized by a regime of freely moving cells. During this phase, in fact, tumor cells have not yet formed a macroscopically visible structure, resulting in interactions between tumor cells and the immune system that predominantly occur at the cellular level. This phase is particularly critical, as the competition between tumor cells and the immune system retains the potential to limit tumor growth.

In this work, we propose a novel kinetic mathematical framework for describing different kinds of spatial interactions between tumors and the immune system.  Building on the models presented in \cite{conte2018qualitative}, which analyze the interactions between tumor and immune cells in a spatially homogeneous context, we introduce two distinct kinetic models that account for the interactions among three populations: tumor cells, active immune cells, and passive immune cells. These dynamics are mediated by the presence of the host environment (healthy cells) and an additional population of interleukins \cite{briukhovetska2021interleukins}, which can enhance the immune response. First, we describe a conservative scenario in which immune cells do not undergo proliferation or apoptosis but instead change their activity state (between active and passive) through interactions with the tumor population or the surrounding environment. Second, we consider a proliferative scenario where immune cells experience both proliferation and death. In these settings, the evolution of the various cellular populations involves key parameters derived from the actual collision frequencies and probability distributions that characterize the microscopic interactions. From the kinetic models, we derive corresponding macroscopic systems in different asymptotic regimes, obtaining various types of diffusive operators that reflect the distinct migratory behaviors of cells. Precisely, both kinetic models may yield a system featuring linear diffusion for three populations. In the conservative scenario, we also derive a model with nonlinear cross-diffusion for the immune population, while the proliferative scenario also leads to a model characterized by nonlinear self-diffusion for the same population. These macroscopic settings are then analyzed within the established framework of dynamical systems theory \cite{guckenheimer2013nonlinear}, aiming to uncover connections between the characteristic features of the different macroscopic models arising from the same kinetic framework. Finally, extensive numerical simulations of the different spatially distributed models are conducted to test the analytical predictions. 

The organization of the paper is as follows. Section \ref{sec:balEq} contains a formal discussion of the collisional and turning operators involved in the kinetic balance equations for the different cell populations, followed by tailoring the general framework to address the specific scenarios of conservative and proliferative immune systems. Section \ref{deri_macroSys} presents the derivation of macroscopic systems characterized by linear diffusion, nonlinear cross-diffusion, and nonlinear self-diffusion, along with a specification of the assumptions made in each case. In Section \ref{Homog_section}, we conduct a qualitative analysis of the spatially homogeneous counterparts of the developed models, examining and comparing equilibrium configurations, stability, and bifurcations through both analytical and numerical methods. Section \ref{Macro_sim} focuses on numerical simulations of the macroscopic spatially distributed models, with the goal of comparing the qualitative evolution of the different populations. Finally, Section \ref{SecConc} discusses the main outcomes of our model and offers perspectives for future research.

\section{Description of the kinetic transport equations}\label{sec:balEq}
We introduce here a general kinetic framework to describe the interactions between tumor cells and immune system cells, aiming at understanding how different spatial behaviors can emerge from distinct interaction mechanisms. Our model encompasses five cell populations: three dynamic populations —tumor cells, active immune cells (which can effectively kill tumor cells), and passive immune cells— and two static components -host environment cells and interleukins, which are specific cytokines that enhance immune system activity \cite{briukhovetska2021interleukins}. The latter serves as a fixed background in our analysis.

Building on the framework introduced in \cite{conte2018qualitative}, we aim to provide a comprehensive description of the system's state by examining the distribution functions $f_i(t,\x,\bv,u)$, indexed by $i$ for the different cell types: tumor cells ($i=1$), active immune system ($i=2$), and passive immune system ($i=3$). At time $t>0$, each cell is characterized by its spatial position $\x\in\Omega\subseteq\RR^n$, its microscopic velocity $\bv\in\bV\subset \RR^n$ (without loss of generality, we assume that $|\bV|=1$) and its internal state variable (referred to as {\it activity}) $u\in\Ub=[-1,1]$. The pair $\{\x,\bv\}\in\Omega\times\bV$ represents the microscopic mechanical state, while $u\in\Ub$ represents the microscopic  biological state. Specifically, this activity variable signifies the cell population's specific capabilities, such as proliferation and invasion for aggressive tumor cells, and defense mechanisms for immune cells. It quantifies a cell's effectiveness during interactions with other cell types. We assume that only binary interactions significantly influence the system's evolution. From these smooth, non-negative distribution functions, we can derive the cellular densities $n_i$ as moments of the distribution functions, expressed as follows:
\begin{equation}\label{macro_dens}
    n_i(t,\x)=\int\limits_\bV \int\limits_\Ub f_i(t,\x,\bv,u) du d\bv.
\end{equation}
In particular, the active immune cells are characterized by a positive state $u\in[0,1]$, while passive immune cells have a negative state $u\in[-1,0]$. As a consequence, the distribution function $f_2(t,\x,\bv,u)$ for the active immune system is defined as 
\begin{equation*}
f_2(t,\x,\bv,u)=
    \begin{sistem}
        0 \qquad\qquad\quad\,\,\,\, u\in[-1,0]\\[0.3cm]
        \tilde{f}_2(t,\x,\bv,u) \quad u\in[0,1],
    \end{sistem}
\end{equation*}
while the distribution function $f_3(t,\x,\bv,u)$ for the passive immune system reads
\begin{equation*}
f_3(t,\x,\bv,u)=
    \begin{sistem}
        \tilde{f}_3(t,\x,\bv,u) \quad u\in[-1,0]\\[0.3cm]
        0 \qquad\qquad\quad\,\,\,\, u\in[0,1].
    \end{sistem}
\end{equation*}
The two background populations remain static over time and are represented by space-dependent density functions $n_4(\x)$ and $n_5(\x)$, indicating host environment cells and interleukins, respectively, with ${\x\in\Omega\subseteq\RR^n}$. The evolution of the system bears a strong resemblance to models and methods from gas kinetic theory, in which interactions among molecules or particles are analogous to mechanical collisions, and the state variable typically represents internal energy. In our framework, the distribution functions $f_i(t,\x,\bv,u)$ evolve according to the following kinetic transport equation:
\begin{equation}\label{gen_trans_eq}
\dfrac{\partial}{\partial t}f_i(t,\x,\bv,u)+\bv\cdot\nabla_\x f_i(t,\x,\bv,u)=\mathcal{L} [f_i](t,\x,\bv,u)+\sum_{j=1}^5Q_{ij}(t,\x,\bv,u)+\sum_{\substack{j,k=1 \\ j\ne i}}^5 J^i_{jk}(t,\x,\bv,u)+\mathcal{K}[f_i](t,\x,\bv,u)\,.
\end{equation}
In this context, the operators have the following interpretations:
\begin{itemize}
\item[$\bullet$] $\mathcal{L} [f_i]$ represents the turning operator modeling the changes in velocity in the $i$-th population;
\item[$\bullet$] $Q_{ij}$ represents the collision operator that takes into account the effect on the $i$-th population of the binary interactions with the $j$-th population (which may be either evolving or background);
\item[$\bullet$] $J^i_{jk}$ represents the collision operator that captures the contributions to the $i$-th population from the interactions between the $j$-th and $k$-th populations (also either evolving or background);
\item[$\bullet$] $\mathcal{K}[f_i]$ collects all contributions to the $i$-th population (if any) that arise from factors other than cellular interactions.
\end{itemize}
\begin{oss*}
       Here, we assume that the processes that determine changes in cell velocity $\bv$ do not affect cell activity $u$ and vice-versa \cite{bellomo2004class}.  
\end{oss*}

Following \cite{othmer2000diffusion,bellomo2004class}, the model propose here features linear transport and stochastic velocity jumps, while binary interactions are stochastically modeled with changes in the biological state $u$ and birth or death mechanisms. 
We first define the operators $Q_{ij}$ and $J^i_{jk}$; the formulation of $\mathcal{K}[f_i]$, if applicable,  will be detailed in the subsequent section on the specific model. A key distinction from gas dynamics is that encounters may not be regulated by conservation laws. In fact, as seen in other fields such as transport theory \cite{duderstadt1982transport}, these interactions can result in the disappearance or proliferation of a participating population without requiring a balance. Following the framework established in \cite{conte2018qualitative}, we derive integral operators of Boltzmann type by employing an equivalent probabilistic formulation \cite{boffi1990equivalence}, utilizing appropriate interaction probabilities per time unit—specifically, collision frequencies and interaction kernels. Each cell of the $i$-th population, characterized by velocity $\bv$ and activity $u$, is represented by the triplet $(i,\bv,u)$. We define $d_{ij}(\bv_*,\bv', u_*,u')$ the collision frequency for encounters, between an $(i,\bv_*,u_*)$ cell and a $(j,\bv',u')$ cell, that are destructive to the $i$-th population. Conversely, $\mu_{ij}(\bv_*,\bv', u_*,u')$ denotes the collision frequency for encounters that promote proliferation in the $i$-th population. For proliferative encounters, the expected density of new $i$-th  cells with activity $u$ and velocity $\bv$ is denoted by $\epsilon_{ij}(\bv_*,\bv',u_*,u';\bv, u)$. The integral

$$
m_{ij}(\bv_*,\bv',u_*,u')=\int\limits_\bV \int\limits_\Ub\epsilon_{ij}(\bv_*,\bv, u_*,u';\bv,u)du\,d\bv
$$
provides the average number of $i$-th cells generated during a proliferative encounter between $(i,\bv_*,u_*)$ and $(j,\bv',u')$ cells. To simplify the notation in the expressions for $Q_{ij}$ and $J^i_{jk}$, we assume the existence of a distribution function $f_j$, for $j=4,5$, such that $n_j(t,\x)=\int_\bV \int_\Ub f_j(t,\x,\bv,u)$ and $n_j(t,\x)=n_j(\x)$. Thus, the kinetic collision operator $Q_{ij}$ can be expressed as follows: 
\begin{equation}\label{coll_ope}
\begin{aligned}
Q_{ij}(t,\x,\bv,u)=&\iint\limits_\bV\iint\limits_\Ub\mu_{ij}(\bv_*,\bv', u_*,u')\epsilon_{ij}(,\bv_*,\bv',u_*,u';\bv, u)f_i(t,\x,\bv_*,u_*)f_j(t,\x,\bv',u')du_*\,du'\,d\bv_*\,d\bv'\\[2mm]
    & -f_i(t,\x,\bv,u)\int\limits_\bV\int\limits_\Ub d_{ij}(\bv,\bv', u,u')
    f_j(t,\x,\bv',u')du'\,d\bv\,.
\end{aligned}
\end{equation}

Concerning the definition of $J^i_{jk}$ and considering an interaction between $(j,\bv_*,u_*)$ and $(k,\bv',u')$ cells that affects the $i$-th population, we introduce $\alpha^i_{jk}(\bv_*,\bv', u_*,u')$, which represents the collision frequency for interactions between $(j,\bv_*,u_*)$ and $(k,\bv',u')$. Additionally, we define $\nu^i_{jk}(,\bv_*,\bv',u_*,u';\bv, u)$ as the expected density of newly generated $(i,\bv,u)$ cells resulting from these interactions. Consequently,
\begin{equation}\label{beta}
   \beta^i_{jk}(\bv_*,\bv',u_*,u')=\int\limits\limits_\bV \int\limits\limits_\Ub \nu^i_{jk}(\bv_*,\bv', u_*,u';\bv,u)du\,d\bv 
\end{equation}
provides the average number of $i$-th  cells generated during such encounters. 
With this notation established, the interactive operator $J^i_{jk}$ can be expressed as
\begin{equation}\label{interc_ope}
J^i_{jk}(t,\x,\bv,u)=\iint\limits_\bV\iint\limits_\Ub\alpha^i_{jk}(\bv_*,\bv', u_*,u')\nu^i_{kj}(\bv_*,\bv',u_*,u';\bv, u)f_j(t,\x,\bv_*,u_*)f_k(t,\x,\bv',u')du_*\,du'\,d\bv_*\,d\bv'\,.
\end{equation}

Concerning the changes in velocity that are not due to free-particle transport in the $i$-th population, we assume that cells exhibit a {\it run-and-tumble} behavior, characterized by alternating periods of straight-line movement (runs) and random (or biased) reorientations (tumble). This specific dynamics is commonly modeled as a scattering process of the microscopic velocity, described by a {\it velocity jump process} \cite{stroock1974some,alt1980biased}. This process is characterized by a turning frequency $\lambda(\x,\bv)$ and a transition probability $T(\x,\bv|\bv')$. The general form of the turning operator that implements this velocity jump process at a kinetic level is expressed as follows:

$$
\mathcal{L}[f_i](t,\x\,\bv,u):=\int\limits_\bV \lambda(\x,\bv') T(\x,\bv|\bv')f_i(t,\x,\bv',u)d\bv'-\int\limits_\bV \lambda(\x,\bv) T(\x,\bv''|\bv)f_i(t,\x,\bv,u)d\bv'',
$$
where $\bv'$ represents the pre-tumbling velocity contributing to the gain term of $\mathcal{L}[f_i]$, while $\bv''$ denotes the post-tumbling velocity involved in the loss term. The transition probability $T(\x,\bv|\bv')\ge 0$ is commonly referred to as the turning kernel. It quantifies the probability that a cell at $\x$ with velocity $\bv'$ will exhibit velocity $\bv$ following a reorientation. As a probability measure over $\bV$, it satisfies the normalization condition:
\begin{equation}\label{T_prop}
 \int\limits_\bV T(\x,\bv|\bv')d\bv=1\,,\qquad \forall\,\, \bv'\in\bV\,,\,\,\forall\,\, \x\in\Omega.
\end{equation}
Various expressions for the transition probability $T(\x,\bv|\bv')$ have been proposed and studied in the literature \cite{ciarletta2016diffusion,othmer2000diffusion,painter2018random}. For instance, when modeling the space of directions as a sphere of constant radius, i.e., $\bV=s\mathbb{S}^{n-1}$, we can assume that each cell has an equal probability of moving in any direction while traveling at a constant speed $s$. This leads to the simplest turning kernel $T(\x,\bv|\bv'):= 1/|\bV|$ which characterizes the so-called {\it Pearson walk} \cite{ciarletta2016diffusion}. In many biological contexts, particularly regarding tumor cell migration, a more natural approach involves making turning decisions based on the cell's environment \cite{loy2020kinetic,conte2022multi,conte2023non}. For example, glioma cells diffuse through the brain by utilizing white matter tracts as highways for their movement \cite{ware2003molecular,giese1996migration,engwer2015glioma,conte2020glioma}. This type of cell migration behavior is known as the {\it contact guidance} phenomenon \cite{han2016oriented}. In such cases, the turning kernel incorporates information about the underlying anisotropy of the tissue through a directional distribution function. Precisely, let denote by $\theta\in\mathbb{S}^{n-1}$ the direction of the velocity, the turning kernel can be defined as $T(\x,\theta|\theta'):=q(\x,\theta)\ge 0$ with the normalization condition  ${\int_{\mathbb{S}^{n-1}}q(\x,\theta)\,d\theta=1}$. In this context, at each location $\x$, $q(\x,\theta)$ represents the distribution of the direction $\theta$ of the fibers that constitute the extracellular matrix. Moreover, $q(\x,\theta)$ is assumed to be symmetric with respect to $\theta$, meaning that $q(\x,\theta)$=$q(\x,-\theta)$ for $\theta\in\mathbb{S}^{n-1}$, as fibers do not exhibit natural polarization. For the purposes of this study, consistent with the approaches taken in \cite{chauviere2007amodeling,painter2013mathematical,hillen2006m5,hillen2013transport}, we adopt the assumption that cells have no memory of their previous velocity prior to reorientation, leading to ${T(\x,\bv|\bv') = T(\x,\bv)}$. Additionally, we assume that the turning frequency is independent of the microscopic velocity, i.e., $\lambda=\lambda(\x)$. Let $p_i(t,\x,u)$ denote the density of cells in the $i$-th population with activity state $u$, defined as 

$$
p_i(t,\x,u):=\int\limits_\bV f_i(t,\x,\bv',u)d\bv'\,.
$$
From equation \eqref{macro_dens}, we have 
$$
n_i(t,\x)=\int\limits_\Ub p_i(t,\x,u)du\,.
$$
Consequently, the turning operator can be rewritten as
\begin{equation}\label{turn_op}
\mathcal{L}[f_i](t,\x\,\bv,u):=\lambda(\x)\left[T(\x,\bv)p_i(t,\x,u)-f_i(t,\x,\bv,u)\right]\,,
\end{equation}
where we utilize the property in equation \eqref{T_prop}. Notably, we observe  that $f_i(t,\x,\bv,u)=T(\x,\bv)p_i(t,\x,u)$ nullifies the operator. Additionally, it holds that
\begin{equation*}\label{L_con_zero}
    \int\limits_\bV \mathcal{L}[f_i](t,\x\,\bv,u) d\bv =0\,.
\end{equation*}
Denote $\f$ as the vector with $f_i$ as its $i$-th component and define $$\mathcal{Q}_i[\f]=\sum\limits_{j=1}^5Q_{ij}(t,\x,\bv,u)\, \qquad \text{and}\qquad {\mathcal{J}^i[\f]=\sum\limits_{\displaystyle\substack{j,k=1 \\ j,k\ne i}}^5 J^i_{jk}(t,\x,\bv,u)}\,.$$ Plugging equations \eqref{coll_ope}, \eqref{interc_ope}, and \eqref{turn_op} into equation \eqref{gen_trans_eq}, we obtain the kinetic transport equation for the evolving populations: 
\begin{equation}\label{gen_trans_eq2}
\dfrac{\partial}{\partial t}f_i(t,\x,\bv,u)+\bv\cdot\nabla_\x f_i(t,\x,\bv,u)=(\mathcal{L} [f_i]+\mathcal{Q}_i[\f] +\mathcal{J}^i[\f]+\mathcal{K}[f_i])(t,\x,\bv,u)\,,
\end{equation}
for $i=1,2,3$. In the following, we outline the specific interactions between these populations and the background in two distinct scenarios. In the first scenario, we examine a conservative evolution of the immune system, characterized by cells that neither proliferate nor undergo cell death. Instead, these cells experience variations in their activity, allowing them to transition between active and passive states. In the second scenario, we incorporate processes of immune system proliferation and cell death, resulting in a non-conservative evolution for all evolving populations. This adds a dynamic complexity to the interactions, influencing the overall behavior of the system.

\subsection{The conservative immune system scenario}\label{case1}
In this first scenario, we model the immune system as a conservative population, wherein the total number of immune cells remains constant over time. This implies that immune cells can only alter their internal state $u$, transitioning between active and passive states, without any proliferation or death mechanisms involved. Mathematically, this translates to the condition:
\begin{equation}\label{I_const}
 I(t)=\sum_{i=2}^3 \int\limits_\Omega n_i(t,\x) d\x=\text{const}\,.   
\end{equation}
Changes in the state of immune cells arise from two distinct interactions. When an active immune cell interacts with a tumor cell, it induces tumor cell death while itself transitioning into a passive state. Conversely, when a passive immune cell interacts with the background population of interleukins, it switches back to an active state. These interactions drive the dynamics of the immune system while maintaining a constant total number of immune cells. Regarding the tumor cells, they are subject to non-conservative dynamics. Specifically, tumor cells are eliminated through interactions with active immune cells and proliferate by interacting with either passive immune cells or the host environment's background species. The described mechanisms are quantified by various collision frequencies and transition probabilities. Precisely, for the evolution of the tumor distribution function, we assume that
\begin{equation}\label{Coeff_tumor_sis1}
\begin{array}{c}
d_{12}(\bv_*,\bv',u_*,u')=\bar{d}_{12}\,,\quad\,\mu_{13}(\bv_*,\bv',u_*,u')=\bar{\mu}_{13}\,,\quad\,\mu_{14}(\bv_*,\bv',u_*,u')=\bar{\mu}_{14}\,,
\\[0.4cm]\epsilon_{13} (\bv_*,\bv',u_*,u';\bv,u)=\bar\epsilon_{13}(\bv,u)\,,\quad \,\epsilon_{14}(\bv_*,\bv',u_*,u';\bv,u)=\bar\epsilon_{14}(\bv,u)\,.
\end{array}
\end{equation}
For the evolution of the immune distribution functions, we assume:
\begin{equation*}\label{Coeff_IS_sis1}
\begin{array}{c}
d_{21}(\bv_*,\bv',u_*,u')=\bar{d}_{21}\,,\quad\, d_{35}(\bv_*,\bv',u_*,u')=\bar{d}_{35}\,,\quad\, 
\alpha^2_{35}(\bv_*,\bv',u_*,u')=\bar{\alpha}^2_{35}\,,\quad\, 
\alpha^3_{12}(\bv_*,\bv',u_*,u')=\bar{\alpha}^3_{12}\,,\\[0.4cm]
\nu^2_{35}(\bv_*,\bv',u_*,u';\bv,u)=\bar\nu^2_{35}(\bv,u)\,, \quad \,\nu^3_{12}(\bv_*,\bv',u_*,u';\bv,u)=\bar\nu^3_{12}(\bv,u)\,.
\end{array}
\end{equation*}
Moreover, $\bar\nu^2_{35}(\bv,u)$ takes non-zero values only in $\bV\times[0,1]$, while $\bar\nu^3_{12}(\bv,u)$ is non-zero in $\bV\times[-1,0]$. All other interaction parameters appearing in \eqref{gen_trans_eq} are set to zero. Recalling the general expression \eqref{beta}, to ensure that \eqref{I_const} holds, we require the following conditions:
\begin{equation}\label{cons_assump}
    \bar{\alpha}^2_{35}\beta^2_{35}=\bar{d}_{35}\quad\text{and} \quad\bar{\alpha}^3_{12}\beta^3_{12}=\bar{d}_{21}\,.
\end{equation}
Consequently, for the first scenario, the kinetic model can be expressed as follows:
\begin{equation}\label{case1_Boltz}
    \begin{sistem}
 \dfrac{\partial}{\partial t}f_1(t,\x,\bv,u)+\bv\cdot\nabla_\x f_1(t,\x,\bv,u)=\mathcal{L} [f_1](t,\x,\bv,u)+\left[Q_{12}+Q_{13}+Q_{14}\right](t,\x,\bv,u)\,,\\[0.5cm]    
 \dfrac{\partial}{\partial t}f_2(t,\x,\bv,u)+\bv\cdot\nabla_\x f_2(t,\x,\bv,u)=\mathcal{L} [f_2](t,\x,\bv,u)+\left[Q_{21}+J^2_{35}\right](t,\x,\bv,u)\,,\\[0.5cm]
\dfrac{\partial}{\partial t}f_3(t,\x,\bv,u)+\bv\cdot\nabla_\x f_3(t,\x,\bv,u)=\mathcal{L} [f_3](t,\x,\bv,u)+\left[Q_{35}+J^3_{12}\right](t,\x,\bv,u).
    \end{sistem}
\end{equation}
This model effectively captures the intricate interactions between the immune system and tumor dynamics, offering a framework for understanding their behaviors under different conditions, particularly in scenarios involving immune system homeostasis and tumor proliferation.

\subsection{The proliferative immune system scenario}\label{case2}
In this second scenario, we introduce the dynamics of proliferation and apoptosis within the immune cell population. Specifically, we propose that interactions between active immune cells and tumor cells not only result in the death of tumor cells but also induce apoptosis in the active immune cells themselves. Apoptosis, or programmed cell death, is a crucial regulatory mechanism that helps modulate the immune response \cite{ekert1997apoptosis}. Additionally, we allow active immune cells to proliferate. When an active immune cell interacts with either another active immune cell or a passive immune cell, it may trigger specific intracellular signaling pathways that promote cell division \cite{assarsson2004nk,daneshpour2019modeling}. The resulting offspring cell is generated in a passive state, corresponding to an activity level $u\in[-1,0]$. Furthermore, the immune system has self-regulatory mechanisms to control the population of active immune cells, thereby preventing excessive inflammation. The tumor population remains unchanged in terms of its dynamics compared to the first scenario. We remark that we conceptualize cell proliferation as being triggered by the microenvironment, which promotes cell growth through factors like nutrient supply or cellular signaling. In this regard, the binary interactions are intended to represent simplified, localized events that trigger cell division, either through interactions between cells or with the host environment, and allow us to capture the influence of the microenvironment on the cell proliferation rates \cite{conte2018qualitative,bellomo2008mathematical}. 

All the various processes outlined above are quantified by specific interaction frequencies and transition probabilities. For the tumor distribution function, we maintain the assumptions stated in \eqref{Coeff_tumor_sis1}. In contrast, for the passive and active immune cells, we adopt the following assumptions:
\begin{equation*}
\begin{array}{c}
d_{21}(\bv_*,\bv',u_*,u')=\bar{d}_{21}\,,\quad\, d_{35}(\bv_*,\bv',u,_*u')=\bar{d}_{35}\,,\\[0.4cm] \mu_{32}(\bv_*,\bv',u_*,u')=\bar{\mu}_{32}\,,\quad\,
\alpha^2_{35}(\bv_*,\bv',u_*,u')=\bar{\alpha}^2_{35}\,,\quad\,\alpha^3_{22}(\bv_*,\bv',u_*,u')=\bar{\alpha}^3_{22}\,,\\[0.4cm] 
\epsilon_{32}(\bv_*,\bv',u_*,u';\bv,u)=\bar\epsilon_{32}(\bv,u)\,,\quad\, \nu^2_{35}(\bv_*,\bv',u_*,u';\bv,u)=\bar\nu^2_{35}(\bv,u)\,,\quad \,\nu^3_{22}(\bv_*,\bv',u_*,u';\bv,u)=\bar\nu^3_{22}(\bv,u).
\end{array}
\end{equation*}
Here, $\bar\nu^2_{35}(\bv,u)$ has non-zero values only in $\bV\times[0,1]$, while $\bar\nu^3_{22}(\bv,u)$ is defined in $\bV\times[-1,0]$. All other interaction parameters that appear in the general kinetic model \eqref{gen_trans_eq2} are set to zero. The strength of the self-regulatory mechanism of the active immune system, aiming to maintain an optimal distribution $f^*_2(t,\x,\bv,u)$, is modeled to depend linearly on the instantaneous deviation from this desired distribution, characterized by a constant rate parameter $\kappa_2$. This self-control mechanism is integrated into the model through the following operator:
\[
\mathcal{K}[f_2]=-\kappa_2(f_2(t,\x,\bv,u)-f^*_2(t,\x,\bv,u)).
\]
This operator effectively drives the active immune cell population toward its optimal state, promoting stability and preventing excessive immune responses.
Therefore, in this second scenario, the kinetic model can be expressed as follows:
\begin{equation}\label{case2_Boltz}
    \begin{sistem}
 \dfrac{\partial}{\partial t}f_1(t,\x,\bv,u)+\bv\cdot\nabla_\x f_1(t,\x,\bv,u)=\mathcal{L} [f_1](t,\x,\bv,u)+\left[Q_{12}+Q_{13}+Q_{14}\right](t,\x,\bv,u)\,,\\[0.5cm]    
 \dfrac{\partial}{\partial t}f_2(t,\x,\bv,u)+\bv\cdot\nabla_\x f_2(t,\x,\bv,u)=\mathcal{L} [f_2](t,\x,\bv,u)+\left[Q_{21}+J^2_{35}+\mathcal{K}[f_2]\right](t,\x,\bv,u)\,,\\[0.5cm]
\dfrac{\partial}{\partial t}f_3(t,\x,\bv,u)+\bv\cdot\nabla_\x f_3(t,\x,\bv,u)=\mathcal{L} [f_3](t,\x,\bv,u)+\left[Q_{35}+Q_{32}+J^3_{22}\right](t,\x,\bv,u).
    \end{sistem}
\end{equation}
This formulation accounts for the new dynamics introduced by the proliferation and apoptosis of immune cells, along with the self-regulation mechanism that promotes homeostasis within the immune system.

\section{Derivation of the macroscopic systems}\label{deri_macroSys}
In this section, we derive the macroscopic reaction-diffusion systems that describe the densities of the evolving populations defined in \eqref{macro_dens}. Our approach starts from the kinetic equations presented in \eqref{case1_Boltz} and \eqref{case2_Boltz}, and it assumes different time scales for the various mechanisms that drive cell evolution.
For this derivation, we introduce a small parameter $\varepsilon$ and rescale the modeled operators $\mathcal{L}[f_i]$, $\mathcal{Q}_i[\f]$, $\mathcal{J}^i[\f]$, and $\mathcal{K}[f_i]$ according to various orders of $\varepsilon$. In this framework, we suppose that the dominant dynamics are governed by the run-and-tumble behavior of the cells, while binary collisions between the evolving populations and the backgrounds are assumed to occur with lower frequency. We distinguish between three distinct macroscopic derivations based on various scalings that capture the slow dynamics of the system. These scalings will result in different types of diffusion processes at the macroscopic level: linear diffusion, nonlinear cross-diffusion, and nonlinear self-diffusion. Specifically, linear diffusion will be derived under conditions where all collision operators operate on the same scale of $\varepsilon$, corresponding to the macroscopic time scale. For the case of nonlinear cross-diffusion, we assume that the collision operators in the tumor equation are on the slowest time scale, while those for the two immune cell types operate on an intermediate time scale. Finally, for nonlinear self-diffusion, only the collision operators for the passive immune system are considered on the intermediate time scale, whereas the tumor and active immune cells are placed on the slowest time scale. Throughout this analysis, we use the terms {\it linear} and {\it nonlinear} to specifically denote the linearity or non-linearity of the relationships with respect to cell density. 

\subsection{Linear diffusion case}\label{LinDif}
In the first case, we consider both systems \eqref{case1_Boltz} and \eqref{case2_Boltz} and we derive the corresponding macroscopic reaction-diffusion models featuring linear diffusion for all the three macroscopic populations, i.e., tumor cells, $n_1$, active immune cells $n_2$, and passive immune cells $n_3$. We consider a fixed small parameter $\varepsilon$ to rescale the involved operators. Precisely, the evidence of several experiments suggests that the biological dynamics are of a smaller order with respect to the mechanical one. Thus, the turning operator for cell motion is taken of order $1/\varepsilon$, while all the cellular interactions, described by $\mathcal{Q}_i[\f]$, $\mathcal{J}^i[\f]$, and $\mathcal{K}[f_i]$ are of order $\varepsilon$. Since we are interested also in the effects on the cell densities of the cell-cell interactions, we take the same order of $\varepsilon$ to scale the temporal derivatives, i.e., $\tau=t\varepsilon$. 

Let us consider the kinetic equation \eqref{gen_trans_eq2} along with the assumptions regarding the time scales of the various operators. We can express the rescaled equations for the distribution functions $f_i(t,\x,\bv,u)$ (for $i=1,2,3$) as follows
\begin{equation}\label{gen_rescal_eq}
    \varepsilon\dfrac{\partial}{\partial t}f_i(t,\x,\bv,u)+\bv\cdot\nabla_\x f_i(t,\x,\bv,u)=\dfrac{1}{\varepsilon}\mathcal{L} [f_i](t,\x,\bv,u)+\varepsilon\left[\mathcal{Q}_i[\f] +\mathcal{J}^i[\f]+\mathcal{K}[f_i]\right](t,\x,\bv,u).
\end{equation}
As proposed in the literature \cite{bellomo2004class}, we consider an expansion of the distributions $f_i(t,\x,\bv,u)$, for $i=1,2,3$, in terms of $\varepsilon$:
\begin{equation}\label{f_exp1}
    f_i^{\varepsilon}(t,\x,\bv,u)=g_{i}^{\varepsilon}(t,\x,\bv,u)+\varepsilon h_i^{\varepsilon}(t,\x,\bv,u)+O(\varepsilon^2)\,.
\end{equation}

Without loss of generality, we assume that the mass is concentrated in the leading order of the expansion, such that
$\displaystyle\int\limits_\bV g_{i}^{\varepsilon}d\bv=\displaystyle\int\limits_\bV f_{i}^{\epsilon}d\bv=p_i^{\varepsilon}(t,\x,u)$. Consequently, we have
\begin{equation}\label{int_h_0}
\int\limits_\bV h_i^\varepsilon(t,\x,\bv,u)d\bv=0\,.
\end{equation}
By substituting the expansion \eqref{f_exp1} into \eqref{gen_rescal_eq}, we find that at the order of $\varepsilon^0$:

$$
\mathcal{L}[g_i^{\varepsilon}]=0 \Longleftrightarrow g_{i}^{\varepsilon}(t,\x,\bv,u)= T(\x,\bv)p_i^{\varepsilon}(t,\x,u)\,.
$$
Thus, we can express $f_i^{\varepsilon}(t,\x,\bv,u)$ as:
\begin{equation}\label{f_exp11}
f_i^{\varepsilon}(t,\x,\bv,u)=T(\x,\bv)p_i^{\varepsilon}(t,\x,u)+\varepsilon h_i^\varepsilon (t,\x,\bv,u)+O(\varepsilon^2)\,.   
\end{equation}
Next, we substitute \eqref{f_exp11} into \eqref{gen_rescal_eq} and integrate with respect to $\bv$ over the domain $\bV$ to obtain
\begin{equation}\label{pi_equa}
     \varepsilon\dfrac{\partial}{\partial t}p_i^{\varepsilon}+
     \underbrace{\nabla_\x \cdot \int\limits_\bV\bv\,T\,p_i^{\varepsilon}d\bv}_{\text{(a)}} +\varepsilon\,\nabla_\x \cdot \int\limits_\bV \bv h_i^{\varepsilon}d\bv=\underbrace{\int\limits_\bV
     \mathcal L[h_i^{\varepsilon}]d\bv}_{\text{(b)}}+\varepsilon\int\limits_\bV\left(\mathcal{Q}_i[{T\,\bf p}^\varepsilon] +\mathcal{J}^i[T\,{\bf p}^\varepsilon]+\mathcal{K}[p_i^\varepsilon]\right) d\bv\,+O(\varepsilon^2),
\end{equation}
where ${\bf p}^\varepsilon$ indicates the vector with $p_i^\varepsilon$ as its $i$-th component. From \eqref{int_h_0}, we conclude that the term $(b)$ equals zero. Moreover, assuming that $T$ satisfies the condition
\begin{equation}\label{T_sim}
  \int\limits_\bV T(\x,\bv)\bv d\bv =0\qquad \forall\,\, \x\in \Omega\,, 
\end{equation}
we find that the term $(a)$ is also zero. 
\begin{oss*}
Assumption \eqref{T_sim} is quite reasonable in various biological processes that model cell migration. For example, in the case of the Pearson walk \cite{ciarletta2016diffusion}, where $T(\x,\bv):= 1/|\bV|$, this condition holds true. Similarly, for a directional distribution function represented as $T(\x,\theta):=q(\x,\theta)$, the assumption is satisfied as well. These scenarios demonstrate the versatility of the assumption in accurately capturing the dynamics of cell movement in biological contexts.
\end{oss*}
\noindent By assumptions \eqref{T_sim} and \eqref{int_h_0}, from equation \eqref{pi_equa} we obtain:
\begin{equation}\label{pi_equa_2}
     \dfrac{\partial}{\partial t}p_i^{\varepsilon}+\nabla_\x \cdot \int\limits_\bV\bv h_i^{\varepsilon}d\bv =\int\limits_\bV\left[\mathcal{Q}_i[{T(\x,\bv)\,\bf p}^\varepsilon] +\mathcal{J}^i[{T(\x,\bv)\,\bf p}^\varepsilon]++\mathcal{K}[p_i^\varepsilon]\right] d\bv +O(\varepsilon)\,.
\end{equation}
To derive an explicit expression for the second term on the left-hand side, we multiply \eqref{gen_rescal_eq} by $\bv$ and integrate over the domain $\bV$:
\begin{equation*}\label{vpi_equa_2}
\begin{aligned}
   & \varepsilon^2\dfrac{\partial}{\partial t}\int\limits_\bV \bv h_i^\varepsilon d\bv+\nabla_\x \cdot \left(\int\limits_\bV \bv\otimes \bv T(\x,\bv)d\bv p_i^{\varepsilon} \right)+\varepsilon \nabla_\x \cdot\left( \int\limits_\bV \bv \otimes \bv h_i^{\varepsilon}d\bv\right)
    \\[2mm]
    &=-\int\limits_\bV\lambda(\x)\bv \,h_i^{\varepsilon}d\bv+\varepsilon\int\limits_\bV\bv\,\left(\mathcal{Q}_i[{T\,\bf p}^\varepsilon] +\mathcal{J}^i[T\,{\bf p}^\varepsilon]+\mathcal{K}[p_i^\varepsilon]\right) d\bv\, +O(\varepsilon^2)\,.
\end{aligned}
\end{equation*}
The leading order part simplifies to
\begin{equation}\label{fv_firstorder}
    \nabla_\x \cdot \left[\mathbb{D}_i(\x) p_i^{\varepsilon} \right]=-\int\limits_\bV\bv h_i^{\varepsilon}d\bv,
\end{equation}
where the diffusion operator is defined as
\begin{equation*}\label{Di_def}
\displaystyle \mathbb{D}_i(\x):=\frac{1}{\lambda(\x)}\int\limits_\bV T(\x,\bv)\bv\otimes \bv d\bv\,.   
\end{equation*}
Substituting \eqref{fv_firstorder} back into \eqref{pi_equa_2} gives us:
\begin{equation}\label{pi_equa_3}
     \dfrac{\partial}{\partial t}p_i^{\varepsilon}(t,\x,u)=\nabla_\x \nabla_\x : \left[\mathbb{D}_i(\x) p_i^{\varepsilon}(t,\x,u)\right]+\mathcal{Q}_i[{\bf p}^\varepsilon(t,\x,u)] +\mathcal{J}^i[{\bf p}^\varepsilon(t,\x,u)]+\mathcal{K}[p_i^{\varepsilon}(t,\x,u)] +O(\varepsilon).
\end{equation}
Here, the operator $\nabla_\x \nabla_\x : [\cdot]$ represents the double divergence with respect to  $\x$. Thus, the corresponding term in \eqref{pi_equa_3} can be expressed as:
\[
\nabla_\x \nabla_\x : \left[\mathbb{D}_i(\x) p_i^{\varepsilon}(t,\x,u)\right]=\nabla_\x\cdot \left[\mathbb{D}_i(\x)\nabla_\x p_i^{\varepsilon}(t,\x,u)\right] + \nabla_\x\cdot\left[ p_i^{\varepsilon}(t,\x,u) \nabla_\x\cdot \mathbb{D}_i(\x)\right]\,.
\]
This term is commonly referred to as {\it myopic diffusion} in the literature \cite{belmonte2013modelling}, and it has been extensively applied in models that describe cell migration in heterogeneous environments \cite{kelkel2012multiscale,lorenz2014class,conte2020glioma,conte2022multi}. Notably, it reduces to the classical Laplacian operator when the diffusion coefficients are spatially homogeneous. Taking the formal limit as $\varepsilon\rightarrow 0$ in \eqref{pi_equa_3} and integrating with respect to $u\in \Ub$ we obtain:
\begin{equation*}\label{pi_equa_4}
     \dfrac{\partial}{\partial t}n_i(t,\x)=\nabla_\x \nabla_\x : \left[\mathbb{D}_i(\x) n_i(t,\x)\right]+\int\limits_\Ub\left[\mathcal{Q}_i[{\bf p}(t,\x,u)] +\mathcal{J}^i[{\bf p}(t,\x,u)]+\mathcal{K}[p_i^{\varepsilon}(t,\x,u)]\right]\,du.
\end{equation*}
This equation is general and does not rely on the specific forms of $\mathcal{Q}_i$, $\mathcal{J}^i$, and $\mathcal{K}$. By specifying their expressions as outlined in \eqref{case1_Boltz} and \eqref{case2_Boltz}, and integrating with respect to $u$, we can derive the particular macroscopic systems for the two scenarios: the conservative immune system and the proliferative immune system. Specifically, for the conservative scenario described in \eqref{case1_Boltz}, the resulting macroscopic system with linear diffusion is given by:
\begin{equation}\label{S1_Lin}
\begin{sistem}
     \dfrac{\partial}{\partial t}\,n_1(t,\x)-\nabla_\x \nabla_\x : \left[\mathbb{D}_1(\x) \,n_1(t,\x)\right]=-\bar{d}_{12}\,n_1(t,\x)\,n_2(t,\x)+\bar{\mu}_{13}m_{13}\,n_1(t,\x)\,n_3(t,\x)+\bar{\mu}_{14}m_{14}\,n_1(t,\x)\,n_4(\x)\,,\\[0.5cm]
     \dfrac{\partial}{\partial t}\,n_2(t,\x)-\nabla_\x \nabla_\x : \left[\mathbb{D}_2(\x) \,n_2(t,\x)\right]=-\bar{d}_{21}\,n_1(t,\x)\,n_2(t,\x)+\bar{\alpha}^2_{35}\beta^2_{35}\,n_3(t,\x)\,n_5(\x)\,,\\[0.5cm]
     \dfrac{\partial}{\partial t}\,n_3(t,\x)-\nabla_\x \nabla_\x : \left[\mathbb{D}_3(\x) \,n_3(t,\x)\right]=-\bar{d}_{35}\,n_3(t,\x)\,n_5(\x)+\bar{\alpha}^3_{12}\beta^3_{12}\,n_1(t,\x)\,n_2(t,\x)\,.
     \end{sistem}
\end{equation}
This system models the evolution of tumor, active, and passive immune cells, with the assumption that only tumor cells undergo proliferation and death, while interactions facilitate a switching behavior between the two sub-populations of immune cells. In contrast, for the proliferative scenario described in \eqref{case2_Boltz}, the corresponding macroscopic system with linear diffusion is given by:
\begin{equation}\label{S2_Lin}
\begin{sistem}
     \dfrac{\partial}{\partial t}\,n_1(t,\x)-\nabla_\x \nabla_\x : \left[\mathbb{D}_1(\x) \,n_1(t,\x)\right]=-\bar{d}_{12}\,n_1(t,\x)\,n_2(t,\x)+\bar{\mu}_{13}m_{13}\,n_1(t,\x)\,n_3(t,\x)+\bar{\mu}_{14}m_{14}\,n_1(t,\x)\,n_4(\x)\,,\\[0.5cm]
     \dfrac{\partial}{\partial t}\,n_2(t,\x)-\nabla_\x \nabla_\x : \left[\mathbb{D}_2(\x) \,n_2(t,\x)\right]=-\bar{d}_{21}\,n_1(t,\x)\,n_2(t,\x)+\bar{\alpha}^2_{35}\beta^2_{35}\,n_3(t,\x)n_5(\x)-\kappa_2[\,n_2(t,\x)-n^*_2(t,\x)]\,,\\[0.5cm]
     \dfrac{\partial}{\partial t}\,n_3(t,\x)-\nabla_\x \nabla_\x : \left[\mathbb{D}_3(\x) \,n_3(t,\x)\right]=-\bar{d}_{35}\,n_3(t,\x)n_5(\x)+n_2(t,\x)\left[m_{32}\bar{\mu}_{32}\,n_3(t,\x)+\bar{\alpha}^3_{22}\beta^3_{22}\,n_2(t,\x)\right]\,.
     \end{sistem}
\end{equation}
This system captures the evolution of tumor, active, and passive immune cells, under the assumption that both tumor and immune cells can proliferate and undergo cell death. The switching behavior between the two sub-populations of immune cells is specifically driven by the influence of interleukins that enhance their activity.

\subsection{Nonlinear cross-diffusion case for conservative immune system}\label{CrossDif}
In the second case, we focus on system \eqref{case1_Boltz}, which pertains to the conservative immune system scenario. We assume distinct time scales for the processes driving cell evolution to derive a nonlinear cross-diffusion-reaction model for the macroscopic densities of tumor cells $n_1$  and total immune system population ${N=\,n_2+n_3}$ within an appropriate hydrodynamic limit.
In this context, the dominant phenomenon remains the run-and-tumble behavior of the cells. Consequently, the corresponding turning operator is assigned an order of $1/\varepsilon^2$. In contrast, we assume that cell-to-cell interactions involving tumor cells -those leading to tumor cell proliferation or death- occur more slowly than the mechanisms that govern the immune system, specifically activation and deactivation processes. We particularly assume that the interactions between the tumor and active immune cells affect the tumor population at a slower rate than they affect the immune cells themselves. This implies that the transition of immune cells from an active to a passive state happens more rapidly than tumor cell apoptosis. Therefore, we model the binary collisions responsible for tumor cell death and proliferation as slow processes of order $\varepsilon^2$, while the interactions that alter the activity of immune cells are considered to occur on an intermediate time scale of order $\varepsilon$. Given our interest in the effects of these interactions on tumor cell density, we scale the temporal derivatives by taking $\tau= t\varepsilon^2$.

Considering system \eqref{case1_Boltz} along with the aforementioned assumptions regarding the different time scales, we can express the rescaled equations for the distribution functions $f_i(t,\x,\bv,u)$ for $i=1,2,3$ as follows:
\begin{equation}\label{case1_Boltz_rescale}
    \begin{sistem}
 \varepsilon^2\dfrac{\partial}{\partial t}f_1(t,\x,\bv,u)+\bv\cdot\nabla_\x f_1(t,\x,\bv,u)=\left[\dfrac{1}{\varepsilon^2}\mathcal{L} [f_1]+\varepsilon^2Q_{12}+\varepsilon^2Q_{13}+\varepsilon^2Q_{14}\right](t,\x,\bv,u)\,,\\[0.5cm]    
 \varepsilon^2\dfrac{\partial}{\partial t}f_2(t,\x,\bv,u)+\bv\cdot\nabla_\x f_2(t,\x,\bv,u)=\left[\dfrac{1}{\varepsilon^2}\mathcal{L} [f_2]+\varepsilon Q_{21}+\varepsilon J^2_{35}\right](t,\x,\bv,u)\,,\\[0.5cm]
\varepsilon^2\dfrac{\partial}{\partial t}f_3(t,\x,\bv,u)+\bv\cdot\nabla_\x f_3(t,\x,\bv,u)=\left[\dfrac{1}{\varepsilon^2}\mathcal{L} [f_3]+\varepsilon Q_{35}+\varepsilon J^3_{12}\right](t,\x,\bv,u)\,.
    \end{sistem}
\end{equation}
Since the dominant run-and-tumble process operates at order $\varepsilon^2$, the distributions $f_i$ can be expressed as follows:
\begin{equation}\label{f_exp2}
    f_i^{\varepsilon}(t,\x,\bv,u)=g_{i}^{\varepsilon}(t,\x,\bv,u)+\varepsilon^2 h_i^{\varepsilon}(t,\x,\bv,u)+O(\varepsilon^3)\,,
\end{equation}
where the perturbation $h_i^\varepsilon$ is of order $O(1)$ and satisfies the same conditions outlined in Section \ref{LinDif}. By substituting \eqref{f_exp2} into \eqref{case1_Boltz_rescale}, we obtain, at order $\varepsilon^0$ 
\begin{equation*}
g_i^{\varepsilon}(t,\x,\bv,u)=T(\x,\bv)p_i^{\varepsilon}(t,\x,u) 
\end{equation*}
and, thus,
\begin{equation}\label{f_exp3}
f_i^{\varepsilon}(t,\x,\bv,u)=T(\x,\bv)p_i^{\varepsilon}(t,\x,u)+\varepsilon^2 h_i^\varepsilon (t,\x,\bv,u)+O(\varepsilon^3)\,.
\end{equation}
For $i=1$, we can replicate the scaling procedure described in Section \ref{LinDif}, leading to the following macroscopic equation for the tumor cell density $n_1$
\begin{equation*}\label{cross_Tumor}
     \dfrac{\partial}{\partial t}\,n_1(t,\x)-\nabla_\x \nabla_\x :\left[\mathbb{D}_1(\x) \,n_1(t,\x)\right]=-\bar{d}_{12}\,n_1(t,\x)\,n_2(t,\x)+\bar{\mu}_{13}m_{13}\,n_1(t,\x)\,n_3(t,\x)+\bar{\mu}_{14}m_{14}\,n_1(t,\x)\,n_4(\x).
\end{equation*}
Instead, for $i=2,3$, by substituting \eqref{f_exp2} into the second and third equations of \eqref{case1_Boltz_rescale}, integrating over the domain $\bV$ with respect to $\bv$, and applying the properties \eqref{T_prop} and \eqref{T_sim} of the turning operator, we arrive at the following equation: 
\begin{equation}\label{cross_23_rescale}
     \varepsilon^2\dfrac{\partial}{\partial t}p_i^{\varepsilon}+\varepsilon^2\,\nabla_{\x}\cdot\int\limits_\bV\bv  h_i^{\varepsilon}d\bv =\varepsilon \int\limits_\bV\left(\mathcal{Q}_i[{T(\x,\bv)\,\bf p}^\varepsilon] +\mathcal{J}^i[{T(\x,\bv)\,\bf p}^\varepsilon]\right) d\bv +O(\varepsilon^3).
\end{equation}
At the leading order, equation \eqref{cross_23_rescale} simplifies to

$$
\int\limits_\bV\left(\mathcal{Q}_i[{T(\x,\bv)\,\bf p}^\varepsilon] +\mathcal{J}^i[{T(\x,\bv)\,\bf p}^\varepsilon]\right) d\bv=0\,.
$$
Expanding this condition for $i=2$ and $i=3$, we get the following relationships:

\allowdisplaybreaks
\begingroup
\begin{align*}
i=2\Rightarrow\qquad &0=\int\limits_\bV\left(\mathcal Q_2[{T(\x,\bv)\,\bf p}^\varepsilon]+\mathcal J^2[{T(\x,\bv)\,\bf p}^\varepsilon]\right)d\bv\\[0.3cm]
&\,\,\,=-\bar{d}_{21}\int\limits_\bV T(\x,\bv)p^{\varepsilon}_2(t,\x,u)\int\limits_\bV\int\limits_{\Ub}T(\x,\bv')p^{\varepsilon}_1(t,\x,u')du'\,d\bv'd\bv \\[0.3cm]
& \quad\,\,\,\,+ \bar{\alpha}^2_{35}\int\limits_\bV \bar\nu^2_{35}(\bv,u) \iint\limits_\bV\iint\limits_{\Ub}T(\x,\bv_*)p^{\varepsilon}_3(t,\x,u_*)T(\x,\bv')p_5^\varepsilon(t,\x,u')du_* du' d\bv_*d\bv'd\bv\\[0.3cm]
&\,\,\,=-\bar{d}_{21}n^{\varepsilon}_1(t,\x)p^{\varepsilon}_2(t,\x,u)+\bar{\alpha}^2_{35}\,n^{\varepsilon}_3(t,\x)n_5^\varepsilon(\x) \int\limits_\bV\bar\nu^2_{35}(\bv,u) d\bv,\\[0.5cm]
i=3\Rightarrow\qquad
&0=\int\limits_\bV\left(\mathcal Q_3[{T(\x,\bv)\,\bf p}^\varepsilon]+\mathcal J^3[{T(\x,\bv)\,\bf p}^\varepsilon]\right)d\bv\\[0.3cm]
&\,\,\,=-\bar{d}_{35}\int\limits_\bV T(\x,\bv)p^{\varepsilon}_3(t,\x,u)\int\limits_\bV\int\limits_{\Ub}T(\x,\bv')p_5^\varepsilon(t,\x,u')du'd\bv'd\bv \\[0.3cm]
&\quad\,\,\,\,+\bar{\alpha}^3_{12}\int\limits_\bV\bar\nu^3_{12}(\bv,u) \iint\limits_\bV\iint\limits_{\Ub}T(\x,\bv_*)p^{\varepsilon}_1(t,\x,u_*)T(\x,\bv')p^{\varepsilon}_2(t,\x,u')du_* du'd\bv_*d\bv'd\bv \\[0.3cm]
&\,\,\,=-\bar{d}_{35}p_3^{\varepsilon}(t,\x,u)n_5^\varepsilon(\x)+\bar{\alpha}^3_{12}n^{\varepsilon}_1(t,\x)n^{\varepsilon}_2(t,\x)\int\limits_\bV\bar\nu^3_{12}(\bv,u)d\bv\,.
\end{align*}
\endgroup
In these expressions, we utilized the assumptions made on $n_j$, for $j=5$, and on the associated parameters. Furthermore, applying assumption \eqref{cons_assump} and integrating the derived relationships with respect to $u$,  we obtain
\begin{equation}\label{0order_rel}
n^{\varepsilon}_3(t,\x)=\dfrac{\bar{d}_{21}n^{\varepsilon}_1(t,\x)n^{\varepsilon}_2(t,\x)}{\bar{d}_{35}n_5(\x)}\,.
\end{equation}
Considering the relation $N^{\varepsilon}=n^{\varepsilon}_2+n^{\varepsilon}_3$, we can derive from \eqref{0order_rel}:
\begin{equation}\label{rel_crosdiff}
    \begin{split}
       & n^{\varepsilon}_2(t,\x)=\dfrac{\bar{d}_{35}n_5(\x)}{\bar{d}_{35}n_5(\x)+\bar{d}_{21}n^{\varepsilon}_1(t,\x)}N^{\varepsilon}(t,\x)\,,\\[0.5cm]
      & n^{\varepsilon}_3(t,\x)=\dfrac{\bar{d}_{21}n^{\varepsilon}_1(t,\x)}{\bar{d}_{35}n_5(\x)+\bar{d}_{21}n^{\varepsilon}_1(t,\x)}N^{\varepsilon}(t,\x)\,.
    \end{split}
\end{equation}
Next, by examining the first order of $\varepsilon$ in \eqref{cross_23_rescale}, we obtain:
\begin{equation}\label{cross_pi_equa_2}
\dfrac{\partial}{\partial t}p_i^{\varepsilon}=-\nabla_\x \cdot \int\limits_\bV\bv  h_i^{\varepsilon}d\bv+O(\varepsilon).
\end{equation}
To express the term on the right-hand side, we substitute \eqref{f_exp3} into the second and third equations of \eqref{case1_Boltz_rescale}, multiply by $\bv$, and integrate with respect to $\bv$ over the domain $\bV$:
\begin{equation*}\label{vpi_equa_2cross}
\begin{split}
   & \varepsilon^4\dfrac{\partial}{\partial t}\int\limits_\bV \bv h_i^\varepsilon d\bv+\nabla_\x \cdot \left(\int\limits_\bV \bv\otimes \bv T(\x,\bv)\,p_i^{\varepsilon} d\bv  \right)+\varepsilon^2 \nabla_\x \cdot \left(\int\limits_\bV \bv \otimes \bv h_i^{\varepsilon}d\bv\right)\\[0.3cm]
   &=-\int\limits_\bV\lambda(\x) \bv h_i^{\varepsilon}d\bv+\varepsilon^3\int\limits_\bV\bv \left(\mathcal{Q}_i[{\bf h}^\varepsilon] +\mathcal{J}^i[{\bf h}^\varepsilon]\right) d\bv +O(\varepsilon^4)\,.
\end{split}
\end{equation*}
At the leading order, this reduces to
\begin{equation*}
    \nabla_\x \cdot \left[\mathbb{D}_i(\x) p_i^{\varepsilon} \right]=-\int\limits_\bV\bv h_i^{\varepsilon}d\bv.
\end{equation*}
 By substituting this condition into \eqref{cross_pi_equa_2}, we obtain:
\begin{equation*}\label{cross_pi_equa_3}
     \dfrac{\partial}{\partial t}p_i^{\varepsilon}(t,\x,u)=\nabla_\x \nabla_\x : \left[\mathbb{D}_i(\x) p_i^{\varepsilon}(t,\x,u)\right]+O(\varepsilon),
\end{equation*}
and, by taking the formal limit as $\varepsilon\rightarrow 0$ and integrating with respect to $u\in \Ub$ for $i=2,3$, we arrive at:
\begin{equation*}\label{cross_pi_equa_4}
     \dfrac{\partial}{\partial t}n_i(t,\x)=\nabla_\x \nabla_\x : \left[\mathbb{D}_i(\x) n_i(t,\x)\right]\,.
\end{equation*}
By summing up the two equations for $i=2$ and $i=3$ and applying the relations in \eqref{rel_crosdiff}, we derive the following macroscopic two-component system with cross-diffusion:
\begin{equation}\label{CrosDif_macro}
\begin{sistem}
\begin{split}
     \dfrac{\partial}{\partial t}\,n_1(t,\x)-\nabla_\x \nabla_\x : \left[\mathbb{D}_1(\x) \,n_1(t,\x)\right]=&\dfrac{\,n_1(t,\x)N(t,\x)}{\bar{d}_{35}n_5(\x)+\bar{d}_{21}\,n_1(t,\x)}\Big(\bar{d}_{21}m_{13}\bar{\mu}_{13}\,n_1(t,\x)
     -\bar{d}_{12}\bar{d}_{35}n_5(\x)\Big)\\[0.5cm]
     &+m_{14}\bar{\mu}_{14}\,n_1(t,\x)\,n_4(\x)\,,
     \end{split}
     \\[1.5cm]
     \dfrac{\partial}{\partial t}N(t,\x) -\nabla_\x \nabla_\x : \left[\dfrac{\mathbb{D}_2(\x)\bar d_{35}n_5(\x)+\mathbb{D}_3(\x)\bar d_{21}\,n_1(t,\x)}{\bar{d}_{35}n_5(\x)+\bar{d}_{21}\,n_1(t,\x)} N(t,\x) \right]=0.
     \end{sistem}
\end{equation}
It is essential that $\mathbb{D}_2(\x)\ne\mathbb{D}_3(\x)$, $\forall\, \x \in \Omega$, to introduce the cross-diffusion term in the equation governing the immune system; otherwise, the system would simplify to the linear diffusion case. The resulting model captures the dynamics of tumor cells and the total immune system. Tumor cells evolve in space through linear diffusion while proliferating and undergoing apoptosis according to the defined reaction terms. In contrast, the immune system, treated as a single entity, evolves spatially solely through the cross-diffusion operator.

\subsection{Nonlinear self-diffusion case for proliferative immune system}\label{SelfDif}
As a third case, we consider system \eqref{case2_Boltz}, which models the proliferative scenario for the immune system. We assume distinct time scales for the processes driving cell evolution in order to derive a nonlinear self-diffusion-reaction system for the macroscopic densities of tumor cells $n_1$ and the total immune system $N=n_2+n_3$,  within an appropriate hydrodynamic limit. 

The dominant phenomenon remains the run-and-tumble mechanism, with the corresponding turning operator characterized by order $1/\varepsilon^2$. We assume that the slowest processes include tumor-active immune cell interactions, leading to tumor cell proliferation or death, immune system lysis, and self-regulation mechanisms. Intermediate time scales, represented by order $\varepsilon$, encompass the dynamics of the passive immune system, specifically proliferation through both passive-active and active-active interactions, as well as the boosting effects of interleukins. We particularly emphasize that the impact of interleukins on the active immune cell population occurs at a slower pace than the interactions themselves: in fact, cells require a certain time to adjust to this changes in activity. Consequently, we model the operator $J^2_{35}$ as being of order $\varepsilon^2$, while the interaction term $Q_{35}$ is of order $\varepsilon$. As done before, to scale the temporal derivatives, we adopt the transformation $\tau=t\varepsilon^2$. 

Considering system \eqref{case2_Boltz} and the assumptions regarding the different time scales, we can write the rescaled equations for the distribution functions  $f_i(t,\x,\bv,u)$ for $i=1,2,3$ as follows:
\begin{equation}\label{case2_Boltz_rescale}
    \begin{sistem}
 \varepsilon^2\dfrac{\partial}{\partial t}f_1t(,\x,\bv,u)+\bv\cdot\nabla_\x f_1(t,\x,\bv,u)=\left[\dfrac{1}{\varepsilon^2}\mathcal{L} [f_1]+\varepsilon^2Q_{12}+\varepsilon^2Q_{13}+\varepsilon^2Q_{14}\right](t,\x,\bv,u)\,,\\[0.5cm]    
 \varepsilon^2\dfrac{\partial}{\partial t}f_2(t,\x,\bv,u)+\bv\cdot\nabla_\x f_2(t,\x,\bv,u)=\left[\dfrac{1}{\varepsilon^2}\mathcal{L} [f_2]+\varepsilon^2 Q_{21}+\varepsilon^2 J^2_{35}+\varepsilon^2 \mathcal{K}[f_2]\right](t,\x,\bv,u)\,,\\[0.5cm]
\varepsilon^2\dfrac{\partial}{\partial t}f_3(t,\x,\bv,u)+\bv\cdot\nabla_\x f_3(t,\x,\bv,u)=\left[\dfrac{1}{\varepsilon^2}\mathcal{L} [f_3]+\varepsilon Q_{35}+\varepsilon Q_{32}+\varepsilon J^3_{22}\right](t,\x,\bv,u)\,.
    \end{sistem}
\end{equation}
For $i=1,2$, the scaling described in Section \ref{LinDif} can be replicated, leading to the following equations for $n_1$ and $n_2$
\begin{equation}\label{self_TumorActive}
\begin{sistem}
     \dfrac{\partial}{\partial t}\,n_1(t,\x)-\nabla_\x \nabla_\x : \left[\mathbb{D}_1(\x) \,n_1(t,\x)\right]=-\bar{d}_{12}\,n_1(t,\x)\,n_2(t,\x)+m_{13}\bar{\mu}_{13}\,n_1(t,\x)\,n_3(t,\x)+m_{14}\bar{\mu}_{14}\,n_1(t,\x)\,n_4(\x)\,,\\[0.5cm]
     \dfrac{\partial}{\partial t}\,n_2(t,\x)-\nabla_\x \nabla_\x : \left[\mathbb{D}_2(\x) \,n_2(t,\x)\right]=-\bar{d}_{21}\,n_1(t,\x)\,n_2(t,\x)+\bar{\alpha}^2_{35}\beta^2_{35}\,n_3(t,\x)n_5(\x)-\kappa_2(\,n_2(t,\x)-\,n_2^*(t,\x))\,.
     \end{sistem}
\end{equation}
For $i=3$, using the expansion \eqref{f_exp2} on $f_3$ and integrating the third equation of \eqref{case2_Boltz_rescale} with respect to $\bv$ over the domain $\bV$, along with the properties of the turning operator, we obtain:
\begin{equation}\label{nonlin_3_rescale}
     \varepsilon^2\dfrac{\partial}{\partial t}p_3^\varepsilon+\varepsilon^2\nabla_\x \cdot \int\limits_\bV\bv  h_3^{\varepsilon}d\bv =\varepsilon \int\limits_\bV (\mathcal{Q}_3[T(\x,\bv) {\bf p}^\varepsilon] + \mathcal{J}^3[T(\x,\bv){\bf p}^\varepsilon]) d\bv +O(\varepsilon^3)\,.
\end{equation} 
At the leading order, equation \eqref{nonlin_3_rescale} simplifies to

$$
\int\limits_\bV\left(\mathcal{Q}_3[{T(\x,\bv)\,\bf p}^\varepsilon] +\mathcal{J}^3[{T(\x,\bv)\,\bf p}^\varepsilon]\right) d\bv=0\,.
$$
Expanding this condition yields the following relationships:

\begin{align*}
0&=\int\limits_\bV\left(\mathcal{Q}_3[T(\x,\bv)\,{\bf p}^\varepsilon] +\mathcal{J}^3[T(\x,\bv)\,{\bf p}^\varepsilon]\right) d\bv\\[0.3cm]
&=-\bar{d}_{35}\int\limits_\bV T(\x,\bv)p^{\varepsilon}_3(t,\x,u)\int\limits_\bV \int\limits_{\Ub}T(\x,\bv')p_5(t,\x,u')du'd\bv' d\bv \\[0.3cm]
&\quad\,\,+\bar{\alpha}^3_{22}\int\limits_\bV\bar{\nu}^3_{22}(\bv,u) \iint\limits_\bV\iint\limits_{\Ub}T(\x,\bv_*)p^{\varepsilon}_2(t,\x,u_*)T(\x,\bv') p^{\varepsilon}_2(t,\x,u') du_* du'd\bv_*d\bv'd\bv \\[0.3cm]
&\quad\,\,+\bar{\mu}_{32}\int\limits_\bV\bar{\epsilon}_{32}(\bv,u) \iint\limits_\bV  \iint\limits_{\Ub}T(\x,\bv_*)p^{\varepsilon}_3(t,\x,u_*)T(\x,\bv')p^{\varepsilon}_2(t,\x,u')du_* du'd\bv_*d\bv'd\bv \,.
\end{align*}
This can be expressed as:

$$
0=-\bar{d}_{35}p_3^{\varepsilon}(t,\x,u) n_5(\x) +\bar{\alpha}^3_{22}n^{\varepsilon}_2(t,\x)n^{\varepsilon}_2(t,\x)\int\limits_\bV\bar\nu^3_{22}(\bv,u) d\bv + \bar{\mu}_{32} n^{\varepsilon}_2(t,\x)n^{\varepsilon}_3(t,\x)\int\limits_\bV\bar\epsilon_{32}(\bv,u)d\bv\,,
$$
where we have made use of the assumptions on $n_j$ for $j=5$ and on the involved parameters. Assuming that $\bar{\mu}_{32}\,m_{32}=\bar{\alpha}^3_{22}\beta^3_{22}$ and integrating with respect to $u$, we obtain the following relation:
\begin{equation*}\label{n3_nonlin}
n^{\varepsilon}_3(t,\x)=\dfrac{\bar{\mu}_{32}m_{32}n^{\varepsilon}_2(t,\x)\big(n^{\varepsilon}_2(t,\x)+n^{\varepsilon}_3(t,\x)\big)}{\bar{d}_{35}n_5(\x)}.
\end{equation*}
From the relation $N^{\varepsilon}=n^{\varepsilon}_2+n^{\varepsilon}_3$, we can derive 
\begin{equation}\label{rel_selfdiff}
    \begin{split}
       & n^{\varepsilon}_2(t,\x)=\dfrac{\bar{d}_{35}n_5(\x)}{\bar{d}_{35}n_5(\x)+N^{\varepsilon}(t,\x)\bar{\mu}_{32}m_{32}}N^{\varepsilon}(t,\x)\,,\\[0.5cm]
      & n^{\varepsilon}_3(t,\x)=\dfrac{\bar{\mu}_{32}m_{32}}{\bar{d}_{35}n_5(\x)+N^{\varepsilon}(t,\x)\bar{\mu}_{32}m_{32}}({N^\varepsilon})^2(t,\x).
    \end{split}
\end{equation}
Next, by considering the first order of $\varepsilon$ in \eqref{nonlin_3_rescale}, we obtain:
\begin{equation}\label{nonlin_pi_equa_2}
\dfrac{\partial}{\partial t}p_3^{\varepsilon}=-\nabla_\x \cdot \int\limits_\bV\bv  h_3^{\varepsilon}d\bv+O(\varepsilon).
\end{equation}
To recover the expression of the first term on the right-hand side, we substitute \eqref{f_exp3} into the third equation of \eqref{case2_Boltz_rescale}, multiply it by $\bv$, and integrate with respect to $\bv$ over the domain $\bV$, yielding:
\begin{equation*}\label{vpi_equa_2Nonlin}
\begin{split}
    &\varepsilon^4\dfrac{\partial}{\partial t}\int\limits_\bV \bv h_3^\varepsilon d\bv+\nabla_\x \cdot \left(\int\limits_\bV \bv\otimes \bv T(\x,\bv)\,p_3^{\varepsilon}d\bv  \right)+\varepsilon^2 \nabla_\x \cdot \int\limits_\bV \bv \otimes \bv h_3^{\varepsilon}d\bv\\[0.3cm]
    &=-\int\limits_\bV \lambda(\x)\bv h_3^{\varepsilon}d\bv+\varepsilon^3\int\limits_\bV\bv \left(\mathcal{Q}_3[{\bf h^\varepsilon}] +\mathcal{J}^3[{\bf h^\varepsilon}]\right) d\bv +\varepsilon^4\int\limits_\bV\bv \mathcal{K}[h_3^\varepsilon] d\bv +O(\varepsilon^4)\,.
    \end{split}
\end{equation*}
The leading order of this equation simplifies to
\begin{equation*}
    \nabla_\x \cdot \left[\mathbb{D}_3(\x) p_3^{\varepsilon} \right]=-\int\limits_\bV\bv h_3^{\varepsilon}d\bv\,.
\end{equation*}
Substituting this relation into \eqref{nonlin_pi_equa_2} yields:
\begin{equation*}\label{nonlin_pi_equa_3}
     \dfrac{\partial}{\partial t}p_3^{\varepsilon}(t,\x,u)=\nabla_\x \nabla_\x : \left[\mathbb{D}_3(\x) p_3^{\varepsilon}(t,\x,u)\right]+O(\varepsilon)\,.
\end{equation*}
By taking the formal limit as $\varepsilon\rightarrow 0$ and integrating with respect to $u\in \Ub$, we obtain:
\begin{equation}\label{self_pi_equa_4}
     \dfrac{\partial}{\partial t}\,n_3(t,\x)=\nabla_\x \nabla_\x : \left[\mathbb{D}_3(\x) \,n_3(t,\x)\right]\,.
\end{equation}
By summing up equation \eqref{self_pi_equa_4} with the second equation in \eqref{self_TumorActive} and applying the relations in \eqref{rel_selfdiff}, we derive the following macroscopic two-component system with nonlinear self-diffusion:
\begin{equation}\label{NonLin_macro}
\begin{sistem}
\begin{split}
     &\dfrac{\partial}{\partial t}\,n_1(t,\x)-\nabla_\x \nabla_\x : \left[\mathbb{D}_1(\x) \,n_1(t,\x)\right]=\\[0.3cm]
     &\dfrac{\,n_1(t,\x)N(t,\x)}{\bar{d}_{35}n_5(\x)+\bar{\mu}_{32}m_{32}N(t,\x)}\Big(m_{13}m_{32}\bar{\mu}_{13}\bar{\mu}_{32}N(t,\x)-\bar{d}_{12}\bar{d}_{35}n_5(\x)\Big)+\bar{\mu}_{14}m_{14}\,n_1(t,\x)\,n_4(\x)\,,
\end{split}\\[1.6cm]
     \begin{split}
     &\dfrac{\partial}{\partial t}N(t,\x)-\nabla_\x \nabla_\x : \left[\dfrac{\mathbb{D}_2(\x)\bar{d}_{35}n_5(\x)+\mathbb{D}_3(\x)\bar{\mu}_{32}m_{32}N(t,\x)}{\bar{d}_{35}n_5(\x)+\bar{\mu}_{32}m_{32}N(t,\x)}N(t,\x)\right]=\\[0.5cm]   &\dfrac{N(t,\x)}{\bar{d}_{35}n_5(\x)+N\bar{\mu}_{32}m_{32}}\Big[-\bar{d}_{12}\bar{d}_{35}n_5(\x)\,n_1(t,\x)+\bar{\mu}_{32}m_{32}\bar{\alpha}^2_{35}\beta^2_{35}n_5(\x)N(t,\x)-\kappa_2\bar{d}_{35}n_5(\x)\Big]+\kappa_2\,n_2^*(t,\x)\,.
     \end{split}
     \end{sistem}
\end{equation}
As already outlined in Section \ref{CrossDif}, it is essential that $\mathbb{D}_2(\x)\ne\mathbb{D}_3(\x)$, $\forall\, \x \in \Omega$, to introduce the nonlinear self-diffusion term in the equation governing the immune system. If this condition were not met, the system would be characterized solely by linear diffusion. The resulting framework describes the evolution of both tumor and immune cells. Tumor cells evolve through linear diffusion while undergoing proliferation and apoptosis according to the specified reaction terms. In contrast, the immune system, treated as a singular entity, evolves spatially via the nonlinear self-diffusion operator, also exhibiting proliferation with self-regulation based on the relevant reaction dynamics.

\section{Qualitative analysis of the spatially homogeneous systems}\label{Homog_section}
In this section, we examine the spatially homogeneous versions of the derived macroscopic systems \eqref{S1_Lin}, \eqref{S2_Lin}, \eqref{CrosDif_macro}, and \eqref{NonLin_macro}, focusing on their equilibrium configurations, stability, and possible bifurcation phenomena.
 
\subsection{The conservative immune system scenario}\label{HomoSisAnal_1}
We start by considering the two conservative systems - the one with linear diffusion \eqref{S1_Lin} and the one with nonlinear cross-diffusion \eqref{CrosDif_macro} - derived in Subsection \ref{LinDif} and \ref{CrossDif}, respectively, which preserve the total immune system population. Specifically, we start from the spatially homogeneous formulation corresponding to the system \eqref{S1_Lin}, which is expressed as follows:

\begin{equation*}
\begin{sistem}
     \dot{\,n_1}(t)=-\bar{d}_{12}\,n_1(t)\,n_2(t)+\bar{\mu}_{13}m_{13}\,n_1(t)\,n_3(t)+\bar{\mu}_{14}m_{14}\,n_1(t)\,n_4\,,\\[0.5cm]
     \dot{\,n_2}(t)=-\bar{d}_{21}\,n_1(t)\,n_2(t)+\bar{\alpha}^2_{35}\beta^2_{35}\,n_3(t)n_5\,,\\[0.5cm]
     \dot{\,n_3}(t)=-\bar{d}_{35}\,n_3(t)n_5+\bar{\alpha}^3_{12}\beta^3_{12}\,n_1(t)\,n_2(t)\,.
     \end{sistem}
\end{equation*}
For simplicity, we introduce the notation $\gamma_i:=\bar{\alpha}^i_{jk}\beta^i_{jk}$ for $i=1,2,3$ and for $j,k=1,..,5$, and $\delta_{ij}:=\bar{\mu}_{ij}m_{ij}$, for $i=1,2,3$ and for $j=1,...,5$. From the conservation condition \eqref{cons_assump}, we find that $\gamma_3=\bar{d}_{21}$ and $\gamma_2=\bar{d}_{35}$. Under these assumptions and recalling expression \eqref{I_const}, we can reformulate the spatially homogeneous two-population system for tumor and active immune cells as:

\begin{equation*}
\begin{sistem}
     \dot{\,n_1}(t)=-\bar{d}_{12}\,n_1(t)\,n_2(t)+\delta_{13}\,n_1(t)[I-\,n_2(t)]+\delta_{14}\,n_4\,n_1(t)\,,\\[0.5cm]
     \dot{\,n_2}(t)=-\bar{d}_{21}\,n_1(t)\,n_2(t)+\bar{d}_{35}n_5[I-\,n_2(t)]\,.
     \end{sistem}
\end{equation*}
To analyze this system, we find it beneficial to convert the equations into dimensionless form by normalizing all densities with respect to a typical density, such as $n_4$, and rescaling time based on a characteristic time, which we can take as the inverse of the tumor proliferation rate in the absence of the immune system, i.e. $\tau=(\delta_4 \,n_4)^{-1}$. Letting $Y_i(t):=n_i(t)/\,n_4$, for $i=1,..,5$, denote the dimensionless densities, and using the same symbol for the new time variable, we can rewrite the dimensionless system as:
\begin{equation}\label{Homo_2pop_Lin1}
\begin{sistem}
     \dot{\,Y_1}(t)=-A\,Y_1(t)\,Y_2(t)+B\,Y_1(t)\big[I-\,Y_2(t)\big]+\,Y_1(t)\,,\\[0.5cm]
     \dot{\,Y_2}(t)=-A\,Y_1(t)\,Y_2(t)+C\big[I-\,Y_2(t)\big]\,.
     \end{sistem}
\end{equation}
Here, we assume that $\bar{d}_{12}=\bar{d}_{21}$ and use $I$ for the rescaled parameter $I/\,n_4$. The other parameters are defined as follows:
\begin{itemize}
    \item $A=\dfrac{\bar{d}_{12}}{\delta_{14}}$ represents the rate at which tumor cells are eliminated by active immune cells, which subsequently become passive after this interaction;
    \item $B=\dfrac{\delta_{13}}{\delta_{14}}$ denotes the proliferation rate of tumor cells resulting from interactions with the passive immune system;
    \item $C=\dfrac{\bar{d}_{35}\,Y_5}{\delta_{14}}$ indicates the rate at which passive immune cells interact with interleukins.
\end{itemize}
It can be easily proved that the first quadrant is a positively invariant set, ensuring the positivity of solutions when starting from positive initial conditions. Additionally, the trajectory $Y_1(t)=0$, $\forall\, t>0$, cannot be crossed. The system \eqref{Homo_2pop_Lin1} has the following equilibrium points:
\[
E_1=\left(0,I\right)\qquad\text{and}\qquad E_2=\left(\dfrac{C}{A}\,\dfrac{AI-1}{1+BI},\dfrac{1+BI}{A+B}\right),
\]
where $E_1$ represents the disease-free equilibrium, indicating the optimal state for the organism, while $E_2$ reflects a coexistence scenario. The existence of $E_2$ is only admissible when it lies within the phase space, specifically for $A\ge 1/I$. Local stability properties of the equilibrium states $E_1$ and $E_2$ can be determined by analyzing the eigenvalues of the Jacobian matrix associated with the system \eqref{Homo_2pop_Lin1}. The Jacobian at $E_1$, denoted as $\mathbb{J}(E_1)$, is a lower triangular matrix with diagonal elements $\lambda_1=1-AI$ and $\lambda_2=-C$. Therefore, $E_1$ is locally asymptotically stable if and only if $A>1/I$, which implies the presence of the coexistence equilibrium state $E_2$. For the stability of $E_2$, the Jacobian matrix $\mathbb{J}(E_2)$ is given by
\begin{equation*}
\mathbb{J}(E_2)=
  \begin{pmatrix}  
0 & -C\,\dfrac{(AI-1)\,(A+B)}{A\,(1+BI)}\\[0.5cm]
-A\,\dfrac{1+BI}{A+B} & -C\,I\,\dfrac{A+B}{1+BI}
  \end{pmatrix}
  \,.
\end{equation*}
The determinant and trace of the Jacobian are 

$$det(\mathbb{J}(E_2))=-C(AI-1) \qquad \text{and} \qquad Tr(\mathbb{J}(E_2))=-C\,I\,\dfrac{A+B}{1+BI}<0\,,$$respectively. In the region where $E_2$ is an admissible equilibrium (i.e., $A\ge 1/I$), we observe that $det(\mathbb{J}(E_2))<0$, indicating that $E_2$ is a saddle point whenever it exists. Consistent with the findings in \cite{conte2018qualitative, iori1999analysis}, we can conclude that the bifurcation occurring at $A = 1/I$ is a transcritical bifurcation of the forward type. Additionally, as noted in \cite{conte2018qualitative}, the condition $AI\ge1$, which enables tumor depletion and makes the disease-free equilibrium asymptotically stable, relates to the effectiveness of the immune system's response in controlling tumor growth relative to the tumor's proliferation rate in the absence of immune intervention.

\begin{figure}[ht!]
    \centering
   \includegraphics[scale=0.75]{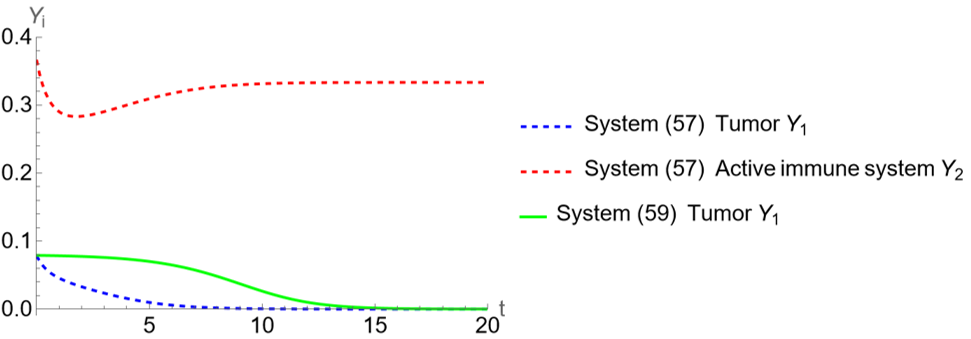}
    \caption{{\bf Qualitative evolution of systems \eqref{Homo_2pop_Lin1} and \eqref{Homo_Cross}.} Qualitative evolution of the trajectories of model \eqref{Homo_2pop_Lin1} (blue dashed line: tumor ($Y_1$), red dashed line: active immune system ($Y_2$)) and \eqref{Homo_Cross} (green solid line: tumor ($Y_1$)) as they approach the equilibrium point $E_1$. The parameter values are set to $A = 5,\,B = 2,\,C = 1,\,I = 0.33$. The trajectories illustrate the dynamics leading to the stable disease-free equilibrium, with differences in transient behaviors observed between the two systems.}
    \label{PlotCons}
\end{figure}

We now turn our attention to the analysis of system \eqref{CrosDif_macro}, which is derived from the same kinetic model \eqref{case1_Boltz} as system \eqref{S1_Lin}, although with a different scaling approach. Notably, the absence of a reaction term for the overall immune system population leads to the following homogeneous formulation:

\begin{equation*}
\begin{sistem}
     \dot{\,n_1}(t)=\dfrac{\,n_1(t)N(t)}{\bar{d}_{35}n_5+\bar{d}_{21}\,n_1(t)}\Big(\bar{d}_{21}m_{13}\bar{\mu}_{13}\,n_1(t)
     -\bar{d}_{12}\bar{d}_{35}n_5\Big)+m_{14}\bar{\mu}_{14}\,n_1(t)\,n_4\,,\\[0.5cm]
     \dot{N}(t)=0.
     \end{sistem}
\end{equation*}
The second equation, combined with the initial condition $N(0)=I$, simplifies to $N(t)=I,\,\,\forall\, t>0$. Therefore, substituting this into the equation for tumor evolution, we obtain:

\begin{equation*}
     \dot{\,n_1}(t)=\dfrac{\,n_1(t)I}{\bar{d}_{35}n_5+\bar{d}_{21}\,n_1(t)}\Big(\bar{d}_{21}\delta_{13}\,n_1(t)
     -\bar{d}_{12}\bar{d}_{35}n_5\Big)+\delta_{14}\,n_1(t)\,n_4\,.
\end{equation*}
Defining the characteristic time $\tau=(\delta_4 \,n_4)^{-1}$ and introducing dimensionless densities $Y_i(t):=n_i(t)/\,n_4$, for ${i=1,..,5}$, the dimensionless equation for tumor cells becomes:
\begin{equation}\label{Homo_Cross}
     \dot{\,Y_1}(t)=\dfrac{A \,I\,Y_1(t)}{C+A\,Y_1(t)}(B\,Y_1(t)-C)+\,Y_1(t)\,,
\end{equation}
where $I$ denotes the rescaled parameter $I/\,n_4$. As in the previous discussion, we can confirm that the positivity of solutions is guaranteed when starting from positive initial conditions. Furthermore, $Y_1(t)=0$, $\forall\, t>0$, is a trajectory. The equilibrium points of equation \eqref{Homo_Cross} are given by:
\[
E_1=0\qquad\text{and}\qquad E_2=\dfrac{C}{A}\,\dfrac{AI-1}{1+BI}
\]
with the condition that $N(t)=I$. Here, $E_1$ represents the disease-free equilibrium, while $E_2$ corresponds to a coexistence scenario that is admissible only when $I\ge 1/A$, indicating sufficient immune system presence. To examine the stability of these equilibria, we define the vector field
\[
f(\,Y_1)=\dfrac{A \,I\,Y_1(t)}{C+A\,Y_1(t)}(B\,Y_1(t)-C)+\,Y_1(t)=\,Y_1\left(1+A\,I\,\dfrac{B\,Y_1-C}{C+A\,Y_1}\right)\,.
\]
Calculating the derivative with respect to $Y_1$ at $E_1$ 

$$\dfrac{\partial\, f(\,Y_1)}{\partial\, \,Y_1}\big|_{\,Y_1=E_1}=1-AI,$$ we find that $E_1$ is locally asymptotically stable if and only if $I\ge 1/A$. To assess the stability of $E_2$, we analyze the sign of the vector field, which can be expressed as:

\begin{equation*}
\begin{split}
 f(\,Y_1)\ge0&\,\Longleftrightarrow\, \dfrac{C\,(1-A\,I)+A\,Y_1(1+B\,I)}{C+A\,Y_1}\ge0 \\[0.3cm]
 &\,\Longleftrightarrow\, \,Y_1\ge\dfrac{C}{A}\,\dfrac{AI-1}{1+BI}=E_2.
 \end{split}
\end{equation*}
Thus, $E_2$ is an unstable equilibrium whenever it exists. The bifurcation occurring at $I = 1/A$ is identified as a transcritical bifurcation of the forward type. This qualitative analysis of the homogeneous counterparts of systems \eqref{S1_Lin} and \eqref{CrosDif_macro} reveals strikingly similar results. Both systems possess two equilibria: a disease-free equilibrium ($E_1$) and a coexistence equilibrium ($E_2$), with $E_1$ exhibiting changes in stability while $E_2$, when present, remains unstable. The conditions for the existence of $E_2$ and the change in stability for $E_1$ relate directly to the immune system's efficacy in fighting the tumor. Therefore, the asymptotic behavior of the modeled conservative system is determined not by the time scales of the various processes involved, but rather by the underlying characteristics of the kinetic model itself. Figure \ref{PlotCons} illustrates the trend toward the stable disease-free equilibrium $E_1$ for both systems: \eqref{Homo_2pop_Lin1} (dashed lines) and \eqref{Homo_Cross} (solid line), using parameter values $A = 5,\,B = 2,\,C = 1,\,I = 0.33$. It is evident that, during the transient phase, tumor depletion occurs more slowly under the conditions imposed by the scaling assumptions discussed in Subsection \ref{CrossDif}.

\subsection{The proliferative immune system scenario}\label{OmoSisAnal_2}
We now turn to the two proliferative settings: one characterized by linear diffusion \eqref{S2_Lin} and the other by nonlinear self-diffusion \eqref{NonLin_macro}, derived in Subsections \ref{LinDif} and \ref{SelfDif}, respectively. In these settings, the proliferation of the passive immune system is taken into account. Starting with system \eqref{S2_Lin} and adopting the previously defined notation, the corresponding homogeneous formulation is given by:

\begin{equation*}
\begin{sistem}
\dot{\,n_1}(t)=-\bar{d}_{12}\,n_1(t)\,n_2(t)+\delta_{13}\,n_1(t)\,n_3(t)+\delta_{14}\,n_1(t)\,n_4\,,\\[0.5cm]
\dot{\,n_2}(t)=-\bar{d}_{21}\,n_1(t)\,n_2(t)+\gamma_2\,n_3(t)n_5-\kappa_2[\,n_2(t)-\,n_2^*]\,,\\[0.5cm]
\dot{\,n_3}(t)=\bar{d}_{35}\,n_3(t)n_5+n_2(t)\left[\delta_{32}\,n_3(t)+\gamma_3\,n_2(t)\right]\,.
\end{sistem}
\end{equation*}
Given the characteristic time $\tau=(\delta_4 \,n_4)^{-1}$ and defining the dimensionless densities $Y_i(t):=n_i(t)/\,n_4$, for $i=1,..,5$, the dimensionless system for tumor cells, active immune cells, and passive immune cells becomes:
\begin{equation}\label{Homo_3pop_Lin2}
\begin{sistem}
     \dot{\,Y_1}(t)=-A\,Y_1(t)\,Y_2(t)+B\,Y_1(t)\,Y_3(t)+\,Y_1(t)\,,\\[0.5cm]
     \dot{\,Y_2}(t)=-A\,Y_1(t)\,Y_2(t)+D\,Y_3(t)-G\,Y_2+H\,,\\[0.5cm]
     \dot{\,Y_3}(t)=-C\,Y_3(t)+E\,Y_2(t)\,Y_3(t)+F\,Y_2^2(t)\,.
     \end{sistem}
\end{equation}
In this formulation, the new parameters are defined as follows:
\begin{itemize}
    \item $D=\dfrac{\gamma_2\,Y_5}{\delta_{14}}$ is the activation rate of immune cells due to interactions between passive immune cells and interleukins;
    \item $E=\dfrac{\delta_{32}}{\delta_{14}}$ is the proliferation rate of passive immune cells due to interactions of active and passive immune cells;
    \item $F=\dfrac{\gamma_3}{\delta_{14}}$ is the proliferation rate of passive immune cells due to interactions between two active immune cells;
    \item $G=\dfrac{\kappa_2}{\delta_{14}\,n_4}$ is the relaxation rate of immune cells towards their optimal distribution $Y_2^*$;
    \item $H=\dfrac{\,n_2^*\kappa_2}{\delta_{14}\,n_4}$ si the rescaled value of the optimal distribution $Y_2^*$.
\end{itemize}

\begin{figure}[h]
\centering
\begin{tabular}{ccc}
&
\\
(A)&&(B)
\\
\includegraphics[width=0.3\textwidth]{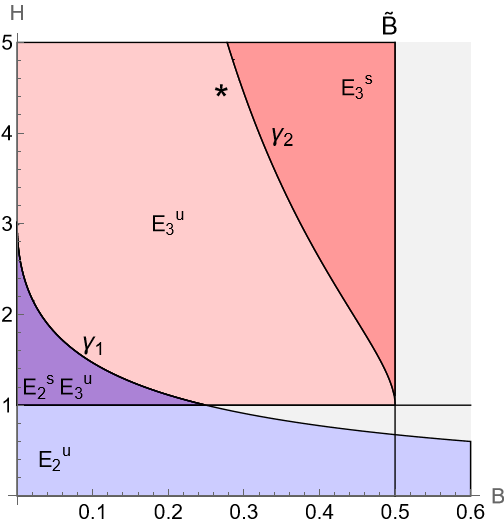} & \qquad\qquad&\includegraphics[width=0.3\textwidth]{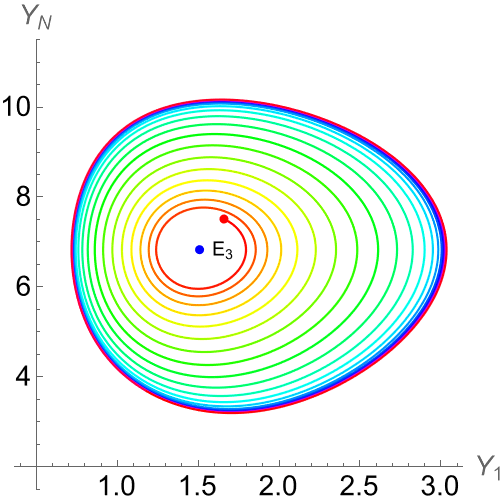} \\
\end{tabular}
\caption{{\bf Bifurcation analysis on system \eqref{Homo_2pop_Nonlin}.} Panel A: bifurcation diagram highlighting existence and stability of equilibria $E_2$ and $E_3$ in system \eqref{Homo_2pop_Nonlin}, setting the parameter values to ${A=G=1.5}$, $P=0.5$, and setting $B=D$. Panel B: Hopf limiting cycle appearing in system \eqref{Homo_2pop_Nonlin} for the parameter values $A=G=1.5$, $P=0.5$, $B=D=0.25$, $H=4.5$. }
\label{Bif_diagram_cycle}
\end{figure}

It can be easily shown that the first octant is a positively invariant set, ensuring that solutions remain positive when initialized with positive conditions. To study the equilibrium points, we assume $E=F$, i.e.,  meaning that the proliferation rate of passive immune cells is independent of the interacting cell types. The system \eqref{Homo_3pop_Lin2} admits two disease-free equilibria: 
\[
E_1=\left(0,{\bar Y_2}^+,\,\frac{G \,{\bar Y_2}^+ -H}{D}\right),\,\quad E_2=\left(0,\,{\bar Y_2}^-,\,\frac{G \,{\bar Y_2}^- -H}{D}\right),
\]
being 

$${\bar Y_2}^{\pm}=\frac{C\,G+E\,H\pm\sqrt{(C\,G-E\,H)^2-4 \,C\,D\,E\,H}}{2\,E (D+G)}.$$
These equilibria are non-negative under the conditions:
\begin{equation*}\label{E1E2Pos}
D\leq\widetilde D:= \frac{(C\,G-E\,H)^2}{4 \,C\,E\,H}\quad\text{and}\quad G>\dfrac{E}{C}\,H\,.
\end{equation*}
Additionally, the system has two coexistence equilibria given by:
\[
E_3=\left(\frac{A\,D-B\,G}{A\,B}+\frac{B\,H-D}{B\,(B\,{\bar Y_3}^++1)},\,\frac{1+B\,{\bar Y_3}^+}{A},\,
{\bar Y_3}^+\right),\quad E_4=\left(\frac{A\,D-B\,G}{A\,B}+\frac{B\,H-D}{B\,(B\,{\bar Y_3}^-+1)},\,\frac{1+B\,{\bar Y_3}^-}{A},\,
{\bar Y_3}^-\right)
\]
with

$$
{\bar Y_3}^{\pm}=\frac{A^2\,C-E\,(A+2\,B)\pm \sqrt{(A\,C-E)^2-4\,B\,C\,E}}{2 B\,E\, (A+B)}.
$$
For these coexistence equilibria to be non-negative, the following conditions must hold:
\begin{equation*}\label{E3E4Pos}
B\leq\widetilde B:= \frac{(A\,C-E)^2}{4 \,C\,E}\,,\quad\,\text{and}\,\quad A>\dfrac{E}{C}\,.
\end{equation*}
These conditions must be considered alongside additional criteria for the parameters $D$ and $E$. However, deriving these conditions analytically, along with those related to stability, is both complex and computationally demanding. Consequently, in line with the findings from Section \ref{HomoSisAnal_1}, which demonstrated a correspondence in equilibrium configurations and stability between systems derived from the same kinetic setting by assuming different scalings, we will examine the homogeneous version of system \eqref{NonLin_macro}. This approach aims to extract insights relevant to system \eqref{Homo_3pop_Lin2}. Using the notation introduced earlier, the homogeneous setting corresponding to system \eqref{NonLin_macro} is expressed as follows:

 \begin{equation*}
 \begin{sistem}
 \dot{\,n_1}(t)=\dfrac{\,n_1(t)N(t)}{\bar{d}_{35}n_5+\delta_{32}N(t)}\Big(\delta_{13}\delta_{32}N(t)-\bar{d}_{12}\bar{d}_{35}n_5\Big)+\delta_{14}\,n_1(t)\,n_4\,,\\[0.3cm]
 \dot{N}(t)=\dfrac{N(t)}{\bar{d}_{35}n_5+\delta_{32}N(t)}\left(-\bar{d}_{12}\bar{d}_{35}n_5\,n_1(t)+\delta_{32}\gamma_2n_5N(t)-k_2\bar{d}_{35}n_5\right)+\kappa_2\,n_2^*\,.
 \end{sistem}
 \end{equation*}
Given the characteristic time $\tau=(\delta_4 \,n_4)^{-1}$ and the dimensionless densities $Y_i(t):=n_i(t)/\,n_4$, for $i=1,..,5$, and $Y_N(t)=N(t)/\,n_4$, the dimensionless system for tumor and total immune cells becomes:
 \begin{equation}\label{Homo_2pop_Nonlin}
 \begin{sistem}
      \dot{\,Y_1}(t)=\dfrac{\,Y_1(t)\,Y_N(t)}{1+P\,Y_N(t)}\Big[B\,P\,Y_N(t)-A\Big]+\,Y_1(t)\,,\\[0.5cm]
      \dot{\,Y_N}(t)=\dfrac{\,Y_N(t)}{1+P\,Y_N(t)}\Big[-A\,Y_1(t)+D\,P\,Y_N(t)-G\Big]+H\,,
      \end{sistem}
 \end{equation}
where we have introduced a new parameter $P$ defined as $P:=\frac{E}{C}$. As for system \eqref{Homo_3pop_Lin2}, this system presents two disease-free equilibria and two coexistence equilibria, expressed as:

$$
E_1=\left(0,\frac{G-H\,P+\sqrt{(G-H\,P)^2-4\,H\,D\,P}}{2\,D\,P}\right),\quad E_2=\left(0,\frac{G-H\,P-\sqrt{(G-H\,P)^2-4\,H\,D\,P}}{2\,D\,P}\right),
$$
$$
E_3=\left(\frac{D\,P\,\left(\bar Y_N^+\right)^2+(H\,P\,-G)\bar Y_N^++H}{A\,\left(\bar Y_N^+\right)^2},\bar Y_N^+\right),\quad E_4=\left(\frac{D\,P\,\left(\bar Y_N^-\right)^2+(H\,P\,-G)\bar Y_N^-+H}{A\,\left(\bar Y_N^-\right)^2},\bar Y_N^-\right)
$$
with
$$
\,Y_N^{\pm}=\frac{A-P\pm\sqrt{(A-P)^2-4\,B\,P}}{2\,B\,P}.
$$
The conditions required for the existence and non-negativity of these equilibria are similar to those identified in the three-population system \eqref{Homo_3pop_Lin2}. Specifically, for $E_1$ and $E_2$ to be non-negative, the following conditions must hold:
\[
D\le \tilde{D}\qquad\text{and}\qquad G>H\,P\,,
\]
For the coexistence equilibria, we summarize the conditions for their non-negativity in Table \ref{Exsi_E3E4}. In all cases, we require $A>P$. Additionally, we define the condition for parameter $B$ as 
\[
B^*:=\dfrac{(AH-G)(G-HP)}{H^2P}\,,
\]
while for the parameter $D$ we define
\[
D^*_\pm:=\dfrac{G-AH}{2P}\left(A-P\pm \sqrt{(A-P)^2-4BP}\right) +BH\,.
\]
\begin{table} [!h]
\begin{center}
   \begin{tabular}{|c|c|c|c|c|} 
   \toprule  
   \rule{0pt}{1.5ex}Scenario &H & B & D& Equilibria \\
  \midrule
  \rule{0pt}{3ex}1&$H<\dfrac{G}{A}$&  $B\le \widetilde B$ & $D^*_-\le D\le D^*_+$ & $E_3$\\[1ex]
  \rule{0pt}{3ex}1.a&&  & $D\ge D^*_+$ & $E_3$, $E_4$\\[1ex] \hline
  \rule{0pt}{4ex}2& $\dfrac{G}{A}\le H <\dfrac{2G}{A+P}$&  $B\le B^*$  & $D< D^*_+$  & $E_4$\\[1ex]
  \rule{0pt}{3ex}2.a&&  & $D\ge D^*_+$ & $E_3$, $E_4$\\[1ex]
  \rule{0pt}{3ex}2.b &&  $B^*<B\le \widetilde B$  & $D^*_-\le D\le D^*_+$ & $E_4$\\[1ex]
  \rule{0pt}{3ex}2.c &&  & $D\ge D^*_+$ & $E_3$, $E_4$\\[1ex] \hline
  \rule{0pt}{4ex}3 &$\dfrac{2G}{A+P}\le H <\dfrac{G}{P}$&  $B\le B^*$  & $D< D^*_+$  & $E_4$\\[1ex]
  \rule{0pt}{3ex}3.a &&  & $D\ge D^*_+$ & $E_3$, $E_4$\\[1ex]
  \rule{0pt}{3ex}3.b &&  $B^*<B\le \widetilde B$  & $\forall\, D>0$ & $E_4$, $E_3$\\[1ex] \hline
  \rule{0pt}{4ex}4 &$ H \ge\dfrac{G}{P}$&  $B\le \widetilde B$ & $\forall\, D>0$ & $E_4$, $E_3$\\[1ex]
\bottomrule
    \end{tabular}
\end{center}
\caption{{\bf Conditions for the coexistence equilibria.} We outline the conditions necessary for the non-negative existence of the equilibrium configurations $E_3$ and $E_4$ in system \eqref{Homo_2pop_Nonlin}, assuming that $A>P$.}
 \label{Exsi_E3E4}
\end{table}
\noindent To investigate the stability of the equilibrium states, we denote the equilibrium points as ${E_k=\left(\bar Y_1^{(k)},\bar Y_N^{(k)}\right)}$. The Jacobian matrix associated with system \eqref{Homo_2pop_Nonlin}, evaluated at $E_k$ is given by 
\begin{equation*}
\mathbb{J}(E_k)=
  \begin{pmatrix}  
\dfrac{\bar Y_N^{(k)}}{1+P\,\bar Y_N^{(k)}}\Big[B\,P\,\bar Y_N^{(k)}-A\Big]+1& \,\,\bar Y_1^{(k)}\left(B-\dfrac{A+B}{(1+P \bar Y_N^{(k)})^2}\right)\\[0.7cm]
\dfrac{-A \bar Y_N^{(k)}}{1+P\,\bar Y_N^{(k)}} & D-\dfrac{D+G+A\bar Y_1^{(k)}}{(1+P\bar Y_N^{(k)})^2} 
  \end{pmatrix}
\end{equation*}
for $k=1,2,3,4$. Our goal is to determine the conditions on the parameters such that this Jacobian matrix satisfies $det(\mathbb{J}(E_k))>0$ and $Tr(\mathbb{J}(E_k))<0$, ensuring that the equilibrium $E_k$ is asymptotically stable. It can be shown that, in the regions of parameter space where they exist, equilibria $E_1$ and $E_4$ are always unstable. To analyze the stability $E_2$ and $E_3$ we focus on a specific scenario where $A=G$ and $B=D$. This assumption is not restrictive for our analysis, as it highlights significant dynamics in the stability of $E_2$ and $E_3$. A more comprehensive stability analysis for systems \eqref{Homo_2pop_Nonlin} and \eqref{Homo_3pop_Lin2} will be addressed in future work. Under the identified conditions, we find:
\begin{itemize}
    \item $E_2$ is stable if $H>1$;
    \item $E_3$ is stable if $H>1$ and $\bar{B}<B<\tilde{B}$, where $\bar{B}=\dfrac{H}{P}\dfrac{(A-P)^2)}{(H+1)^2}$.
\end{itemize}
Figure \ref{Bif_diagram_cycle}-A presents the bifurcation diagrams for the equilibria $E_2$ and $E_3$ as the parameters $B$ and $H$ vary. The diagram delineates regions where one or both equilibria exist, categorizing them as stable (indicated by the superscript $s$) or unstable (indicated by the superscript $u$). Additionally, we highlight the horizontal line corresponding to $H=1$ and the vertical line $B=\tilde B$. The two curves are defined by $\gamma_1: B=\tilde D(H)$, $\gamma_2: B=\bar B(H)$. For this analysis, we have set $A=1.5$ and $P=0.5$. The bifurcation for $E_2$ that occurs at $H=1$ is transcritical bifurcation of forward type. In contrast, for $E_3$, the bifurcation occurring at $B=\bar{B}$ can be identified as a Hopf bifurcation. Specifically, when we consider the values $H=4.5$ and $B=0.5$ (marked by the asterisk in Figure \ref{Bif_diagram_cycle}-A), we observe the emergence of a limit cycle, as illustrated in Figure \ref{Bif_diagram_cycle}-B. In this figure, the blue dot represents the equilibrium $E_3$, while the red dot indicates the initial conditions.

\begin{figure}[ht!]
\centering
\begin{tabular}{cc}
&
\\
(A)&(B)
\\
\includegraphics[scale=0.65]{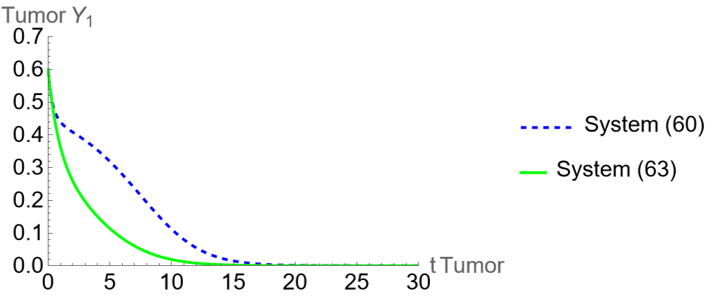} & \includegraphics[scale=0.65]{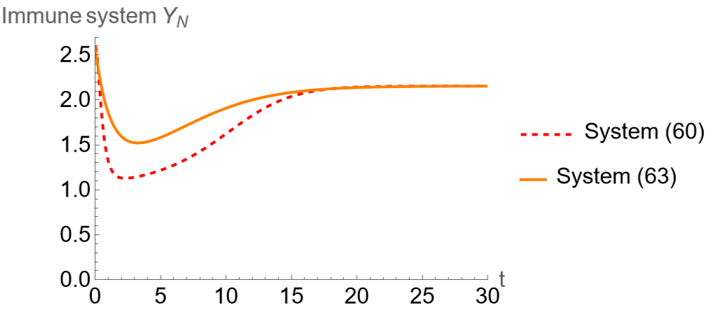} \\
&
\\
(C)&(D)\\
\includegraphics[scale=0.65]{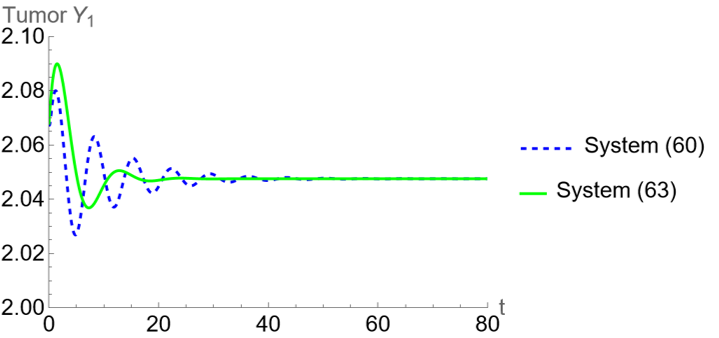} & \includegraphics[scale=0.65]{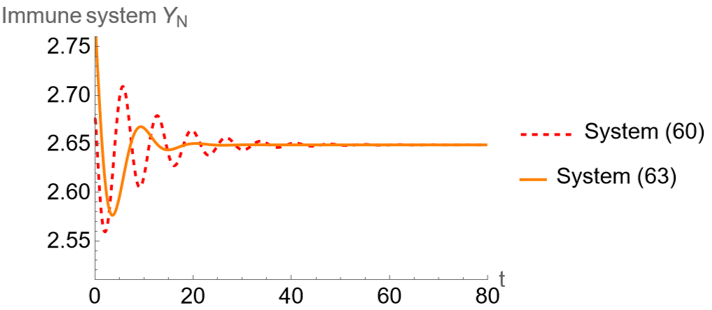} \\
\end{tabular}
    \caption{{\bf Qualitative evolution of systems \eqref{Homo_3pop_Lin2} and \eqref{Homo_2pop_Nonlin}.} Panels A-B: qualitative evolution of the trajectories of model \eqref{Homo_3pop_Lin2} (blue dashed line: tumor ($Y_1$), red dashed line: immune system ($Y_N$)) and \eqref{Homo_2pop_Nonlin} (green solid line: tumor ($Y_1$), orange solid line: immune system ($Y_N$)) as they approach the equilibrium point $E_2$, taking  $B=D=0.05$ and $H=1.5$. Panels C-D: qualitative evolution of the trajectories of model \eqref{Homo_3pop_Lin2} (blue dashed line: tumor ($Y_1$), red dashed line: immune system ($Y_N$)) and \eqref{Homo_2pop_Nonlin} (green solid line: tumor ($Y_1$), orange solid line: immune system ($Y_N$)) as they approach the equilibrium point $E_3$, taking  $B=D=0.47$ and $H=4.5$. In all cases, the remaining parameter values are fixed at $A=1.5$, $C=7$, $E=3.5$, and $P=0.5$.}
    \label{PlotNoncons}
\end{figure}

In line with Section \ref{HomoSisAnal_1}, we expect that the homogeneous counterparts of systems \eqref{S2_Lin} and \eqref{NonLin_macro}, namely \eqref{Homo_3pop_Lin2} and \eqref{Homo_2pop_Nonlin}, will exhibit a significant alignment in their equilibrium stability results. By selecting appropriate parameter values for $B$ and $H$ while fixing $A=1.5$, $C=7$, $E=3.5$, and $P=0.5$, we numerically simulate the behavior of both systems. This allows us to validate the stability results for the non-linear system \eqref{Homo_2pop_Nonlin} and numerically compare them with the dynamics shown by system \eqref{Homo_3pop_Lin2}. Specifically, we first choose $B=D=0.05$ and $H=1.5$, a parameter range where the disease-free equilibrium $E_2$ is stable for system \eqref{Homo_2pop_Nonlin} (panels A-B in Figure \ref{PlotNoncons}). Next, we set $B=D=0.47$ and $H=4.5$, corresponding to the region where the coexistence equilibrium  $E_3$ is stable in the same system (panels C-D in Figure \ref{PlotNoncons}). Specifically, we observe that the stability properties of the equilibria in system \eqref{Homo_2pop_Nonlin} are also confirmed for the non-reduced system \eqref{Homo_3pop_Lin2} (for which the total immune system response $Y_N=\,Y_1+\,Y_2$  is depicted). Additionally, due to the faster timescale associated with the non-conservative processes involving the passive immune system discussed in Subsection \ref{SelfDif}, the equilibrium reached by system \eqref{Homo_2pop_Nonlin} (solid lines) is attained more quickly than for system \eqref{Homo_3pop_Lin2} (dashed lines) in both the disease-free equilibrium (panels A-B) and the coexistence equilibrium (panels C-D).

\section{Numerical simulations of the spatially distributed systems}\label{Macro_sim}
\begin{figure}[h!]
    \centering
    \includegraphics[width=.85\textwidth]{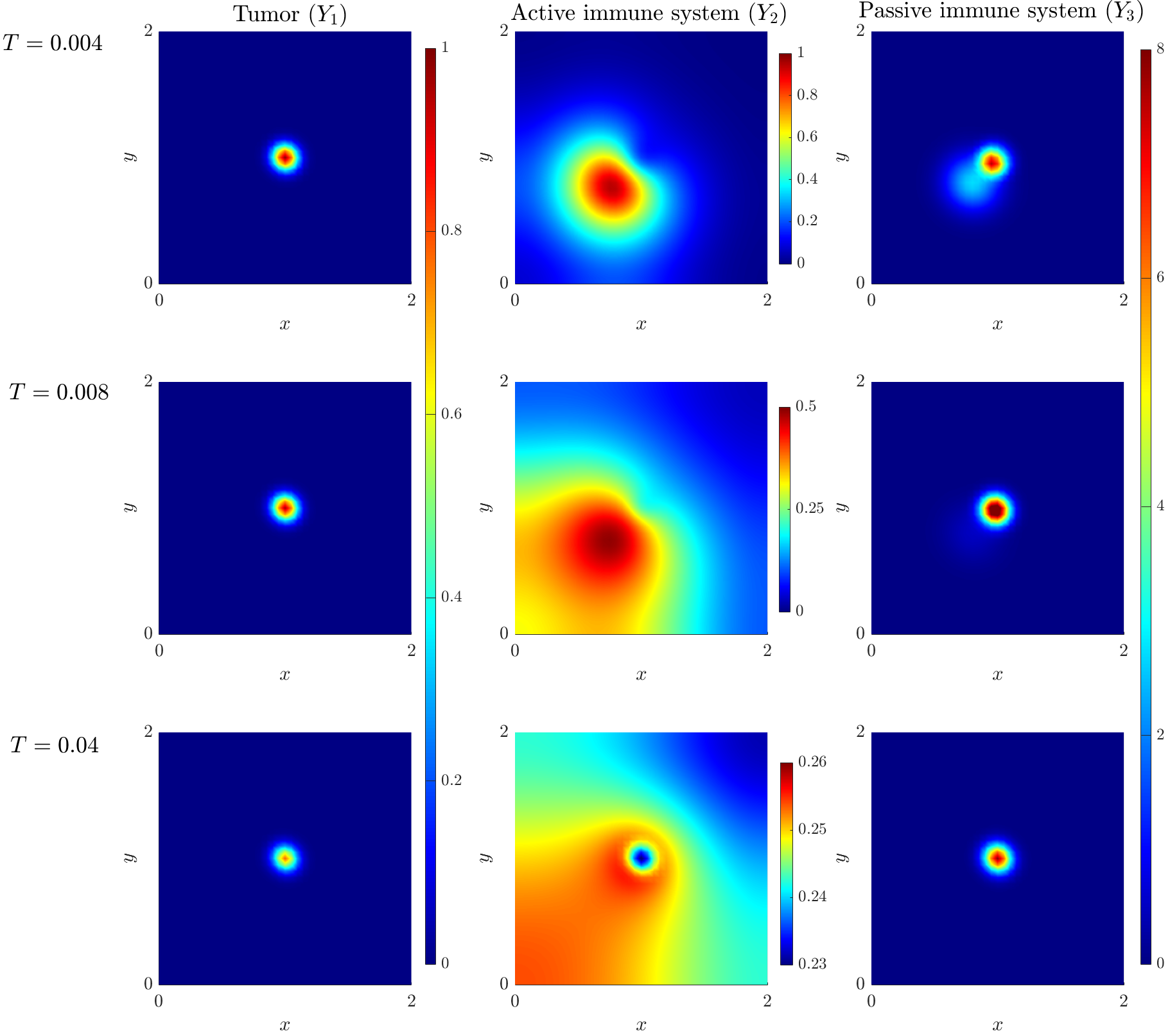}
    \caption{{\bf Test 1: evolution of system \eqref{Adim_Lin1}.} Numerical simulation of model \eqref{Adim_Lin1} with parameter values $A=50$, $B=0.5$, and $C=1$ and the diffusion coefficients $\mathcal{D}_1=10^{-6}$, $\mathcal{D}_2=0.05$, and $\mathcal{D}_3=0.0002$. The columns represent the tumor population ($Y_1$), active immune cells ($Y_2$), and passive immune cells ($Y_3$), respectively. The rows correspond to three different time steps: $T=0.004$, $0.008$, and $0.04$, respectively.}
    \label{2D_Lin1}
\end{figure}

\begin{figure}[h!]
    \centering
   \includegraphics[width=.65\textwidth]{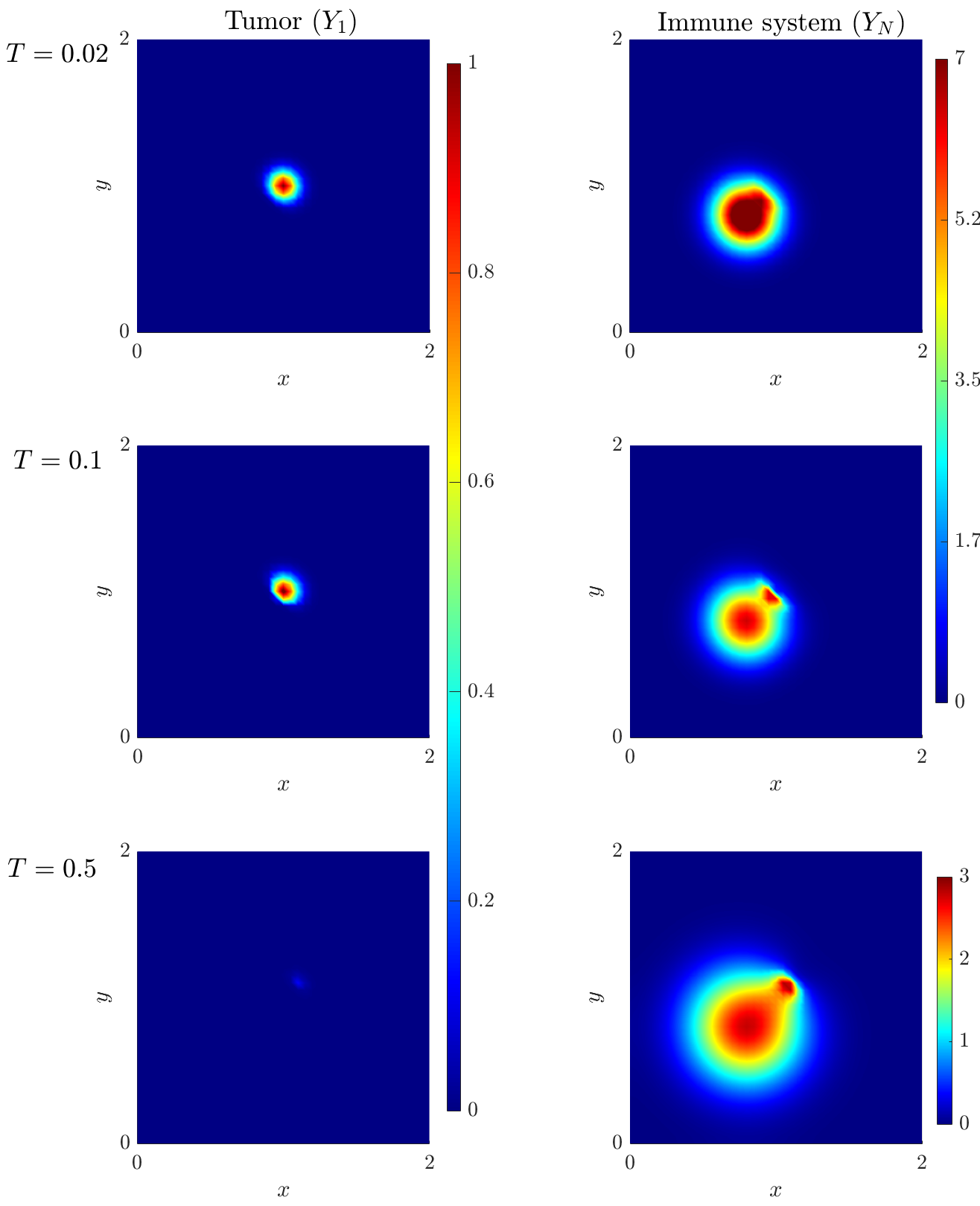}
    \caption{{\bf Test 1: evolution of system \eqref{Adim_Cross}.} Numerical simulation of model \eqref{Adim_Cross} with parameter values $A=50$, $B=0.5$, and $C=1$ and the diffusion coefficients $\mathcal{D}_1=10^{-6}$, $\mathcal{D}_2=0.05$, and $\mathcal{D}_3=0.0002$. The columns represent the tumor population ($Y_1$) and the total immune system ($Y_N$), respectively. The rows correspond to three different time steps: $T=0.02$, $0.1$, and $0.5$, respectively.}
    \label{2D_Cross}
\end{figure}

\begin{figure}[h!]
    \centering
    \includegraphics[width=.45\textwidth]{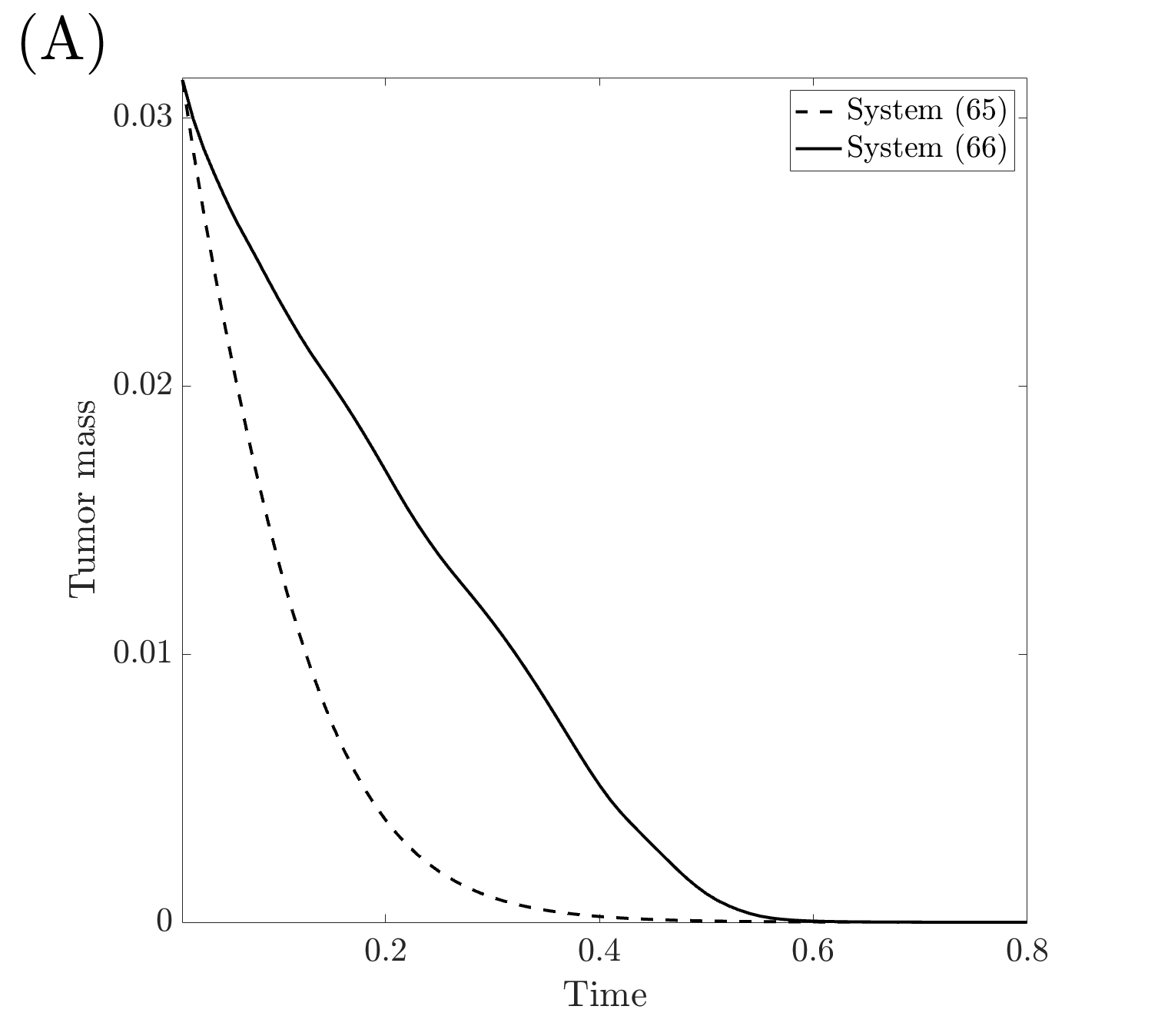}
    \includegraphics[width=.45\textwidth]{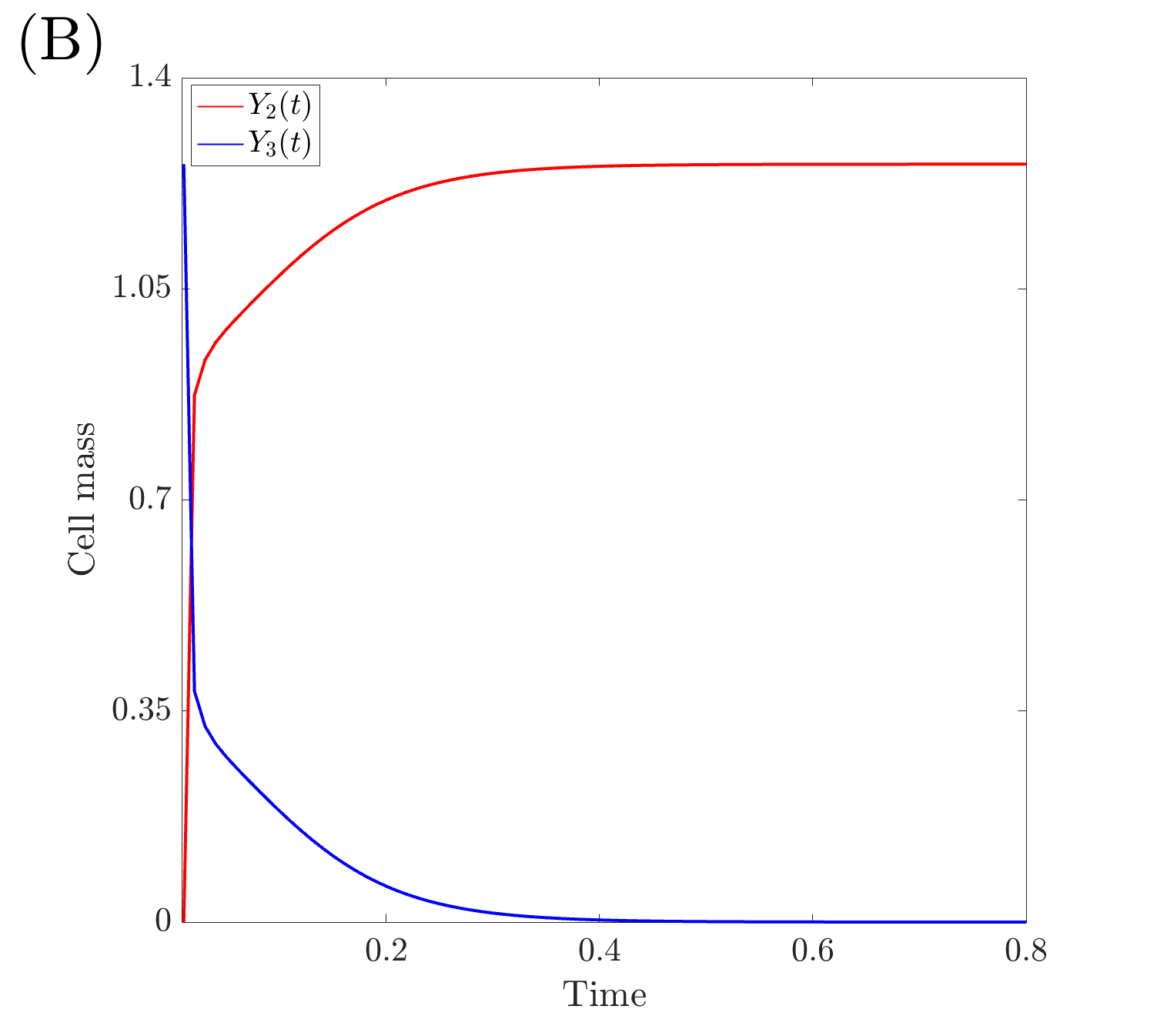}
    \caption{{\bf Test 1: temporal evolution of tumor and immune system masses.} Panel A: comparison of the total tumor mass over time for system \eqref{Adim_Lin1} (black dashed line) and system \eqref{Adim_Cross} (black solid line). Panel B: temporal evolution of the masses of passive (solid blue line) and active (solid red line) immune cells for system  \eqref{Adim_Lin1}. In both cases, the parameter values are set to $A=50$, $B=0.5$, and $C=1$,while the diffusion coefficients are $\mathcal{D}_1=10^{-6}$, $\mathcal{D}_2=0.05$, and $\mathcal{D}_3=0.0002$.}
    \label{Masses_cons}
\end{figure}

We perform 2D numerical simulations of the spatially distributed macroscopic systems \eqref{S1_Lin}, \eqref{S2_Lin}, \eqref{CrosDif_macro}, and \eqref{NonLin_macro}, in their adimensional form. We assume spatially homogeneous diffusion coefficients $\mathcal{D}_i$ for $i=1,2,3$ and uniform distributions for the populations $n_4$ and $n_5$. The simulations are carried out using a self-developed MATLAB code (MathWorks, Inc., Natick, MA). We employ a Galerkin finite element scheme for the spatial discretization of the equations, combined with an implicit Euler scheme for time discretization. Our analysis consists of two primary sets of simulations within the domain $\Omega=[0,2]\times[0,2]$. 
\paragraph{ \bf Test 1: Conservative scenario.} We examine the case of conservative dynamics for the immune population. The adimensionalized version of system \eqref{S1_Lin} is given by:
\begin{equation}\label{Adim_Lin1}
\begin{sistem}
     \dfrac{\partial \,Y_1}{\partial t}(t,\x)-\mathcal{D}_1 \Delta_\x \,Y_1 (t,\x)=-A\,Y_1(t,\x)\,Y_2(t,\x)+B\,Y_1(t,\x)\,Y_3(t,\x)+\,Y_1(t,\x)\,,\\[0.5cm]
     \dfrac{\partial \,Y_2}{\partial t}(t,\x)-\mathcal{D}_2 \Delta_\x \,Y_2(t,\x)=-A\,Y_1(t,\x)\,Y_2(t,\x)+C\,Y_3(t,\x)\,,\\[0.5cm]
      \dfrac{\partial \,Y_3}{\partial t}(t,\x)-\mathcal{D}_3 \Delta_\x \,Y_3(t,\x)=-C\,Y_3(t,\x)+A\,Y_1(t,\x)\,Y_2(t,\x)\,,\\
     \end{sistem}
\end{equation}
while the adimensionalized version of system \eqref{CrosDif_macro} is expressed as:
\begin{equation}\label{Adim_Cross}
\begin{sistem}
      \dfrac{\partial \,Y_1}{\partial t}(t,\x)-\mathcal{D}_1 \Delta_\x \,Y_1(t,\x)=\dfrac{A\,Y_1(t,\x)\,Y_N(t,\x)}{C+A\,Y_1(t)}(B\,Y_1(t)-C)+\,Y_1(t)\,,\\[0.7cm]
      \dfrac{\partial \,Y_N}{\partial t}(t,\x)-\nabla_\x\nabla_\x:\left[\dfrac{\mathcal{D}_2\,C+\mathcal{D}_3A\,Y_1(t,\x)}{C+A\,Y_1(t,\x)}\,Y_N(t,\x)\right]=0\,,\\
         \end{sistem}
\end{equation}
where the parameters $A$, $B$, and $C$ are defined in Section \ref{HomoSisAnal_1}, and $\mathcal{D}_i:=\mathbb{D}_i\,(\delta_{14}\,n_4)^{-1}$ for $i=1,2,3$. For system \eqref{Adim_Lin1}, we consider an initial Gaussian-like aggregate of tumor cells centered at $(x_{0,\,Y_1}, y_{0,\,Y_1})=(1,1)$ with variance $\sigma^2_{\,Y_1}=0.1^{2}$:
\begin{equation}\label{InCon_Tum}
\,Y_{1,0}=\exp\left({-\dfrac{(x-x_{0,\,Y_1})^2+(y-y_{0,\,Y_1})^2}{\sigma^2_{\,Y_1}}}\,\right).
\end{equation}
At the initial stage, we assume the immune system is entirely in a passive state, represented by a Gaussian-like aggregate of passive immune cells centered at $(x_{0,\,Y_3}, y_{0,\,Y_3})=(0.8,0.8)$ with variance $\sigma^2_{\,Y_3}=0.2^{2}$. We set:

$$
\,Y_{2,0}=0\,,\qquad \,Y_{3,0}=\rho_{\,Y_3}\exp\left({-\dfrac{(x-x_{0,\,Y_3})^2+(y-y_{0,\,Y_3})^2}{\sigma^2_{\,Y_3}}}\,\right)
$$
with $\rho_{\,Y_3}=10$. For system \eqref{Adim_Cross}, the initial condition for the entire immune cell population is defined as:

$$
\,Y_{N,0}=\rho_{\,Y_3}\exp\left({-\dfrac{(x-x_{0,\,Y_3})^2+(y-y_{0,\,Y_3})^2}{\sigma^2_{\,Y_3}}}\,\right).
$$
The results of the numerical simulations for models \eqref{Adim_Lin1} and \eqref{Adim_Cross} are illustrated in Figures \ref{2D_Lin1} and \ref{2D_Cross}, respectively.
The dynamics of the tumor populations are qualitatively similar, showing a decrease in tumor mass due to the activity of immune cells. However, the evolution of the immune population differs significantly between the linear diffusion case (Figure \ref{2D_Lin1}) and the nonlinear cross-diffusion case (Figure \ref{2D_Cross}). In particular, the cross-diffusion term for the immune system results in a more confined spread of immune cells within the domain, along with a more pronounced direct movement toward the tumor area, leading to an accumulation of immune cells around the tumor. This effect arises from the nonlinear term in the equation for $Y_N(t,x)$ in \eqref{Adim_Cross}, which can be expressed as:

\begin{equation*}
   \dfrac{\partial \,Y_N}{\partial t}(t,\x)-\nabla_\x\cdot\left[\dfrac{\mathcal{D}_2\,C+\mathcal{D}_3A\,Y_1(t,\x)}{C+A\,Y_1(t,\x)}\,\nabla_\x \,Y_N(t,\x)\right]=\nabla_\x\cdot \left[\,Y_N(t,\x)\nabla_\x\left(\dfrac{\mathcal{D}_2\,C+\mathcal{D}_3A\,Y_1(t,\x)}{C+A\,Y_1(t,\x)}\right)\right]\,.
\end{equation*}
This interaction significantly influences cancer dynamics, resulting in a noticeable asymmetry in the cell configuration. Furthermore, a comparison of the time required for immune cells to eliminate the entire tumor population reveals that it takes longer for the immune system to eradicate the neoplastic mass when cross-diffusion terms are included, as opposed to the linear case. This trend is illustrated in Figure \ref{Masses_cons}-A, which shows the total mass of the tumor population in the domain $\Omega$ over time. 
For completeness, Figure \ref{Masses_cons}-B illustrates the temporal evolution of the total masses of passive and active immune cells in the case of system \eqref{Adim_Lin1}. The plot demonstrates the conservation of the overall immune population, along with the rapid exchange between passive and active states.

\begin{figure}[h!]
    \centering
    \includegraphics[width=.85\textwidth]{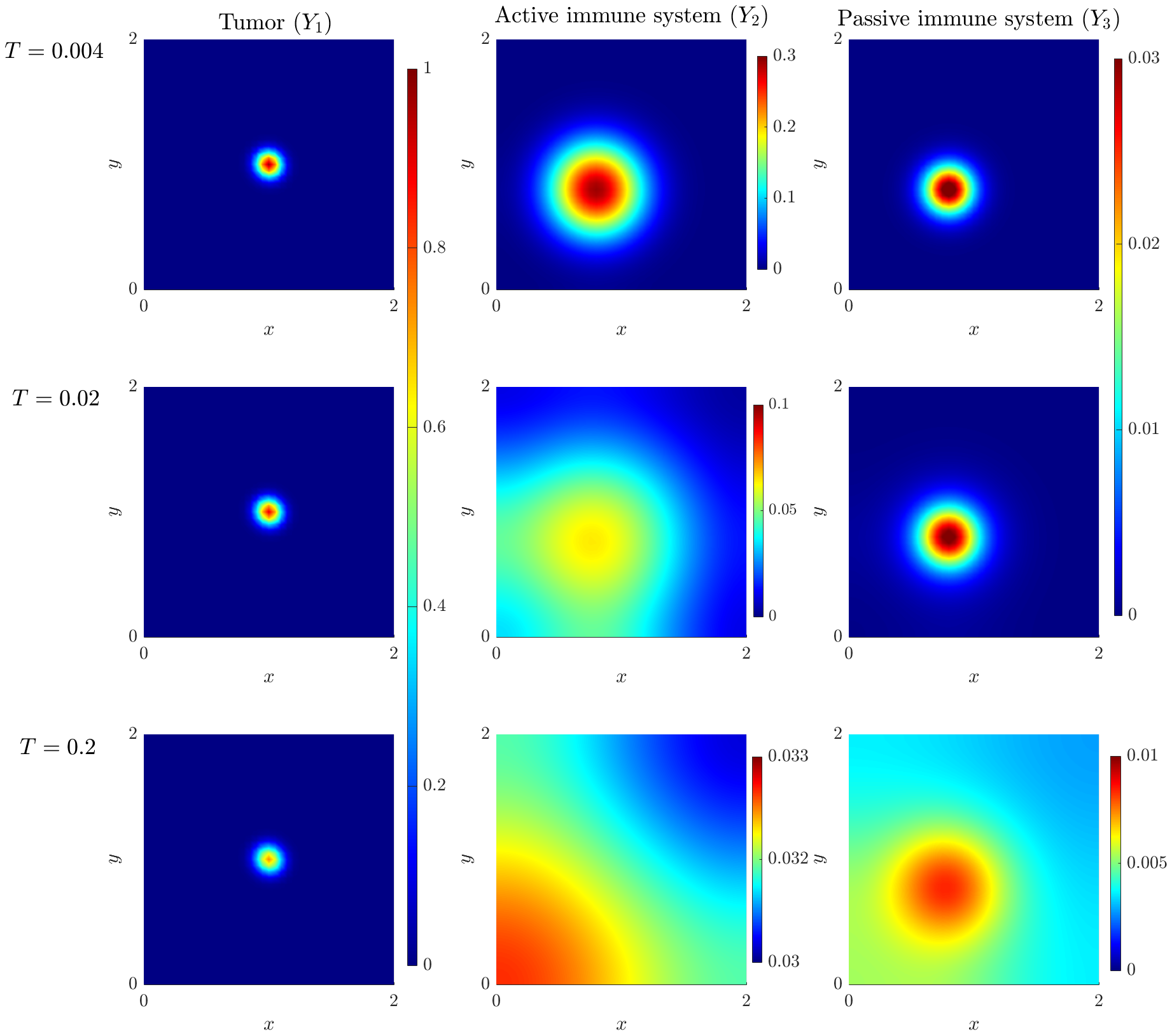}
    \caption{{\bf Test 2: evolution of system \eqref{Adim_Lin2}.} Numerical simulation of model \eqref{Adim_Lin2} with parameter values $A=G=50$, $B=D=1$, $C=7$, $E=F=35$, and $H=1.5$, and the diffusion coefficients $\mathcal{D}_1=10^{-6}$, $\mathcal{D}_2=0.01$, and $\mathcal{D}_3=0.0002$. The columns represent the tumor population ($Y_1$), active immune cells ($Y_2$), and passive immune cells ($Y_3$), respectively. The rows correspond to three different time steps: $T=0.004$, $0.02$, and $0.2$, respectively.}
    \label{2D_Lin2}
\end{figure}

\begin{figure}[h!]
    \centering
   \includegraphics[width=.65\textwidth]{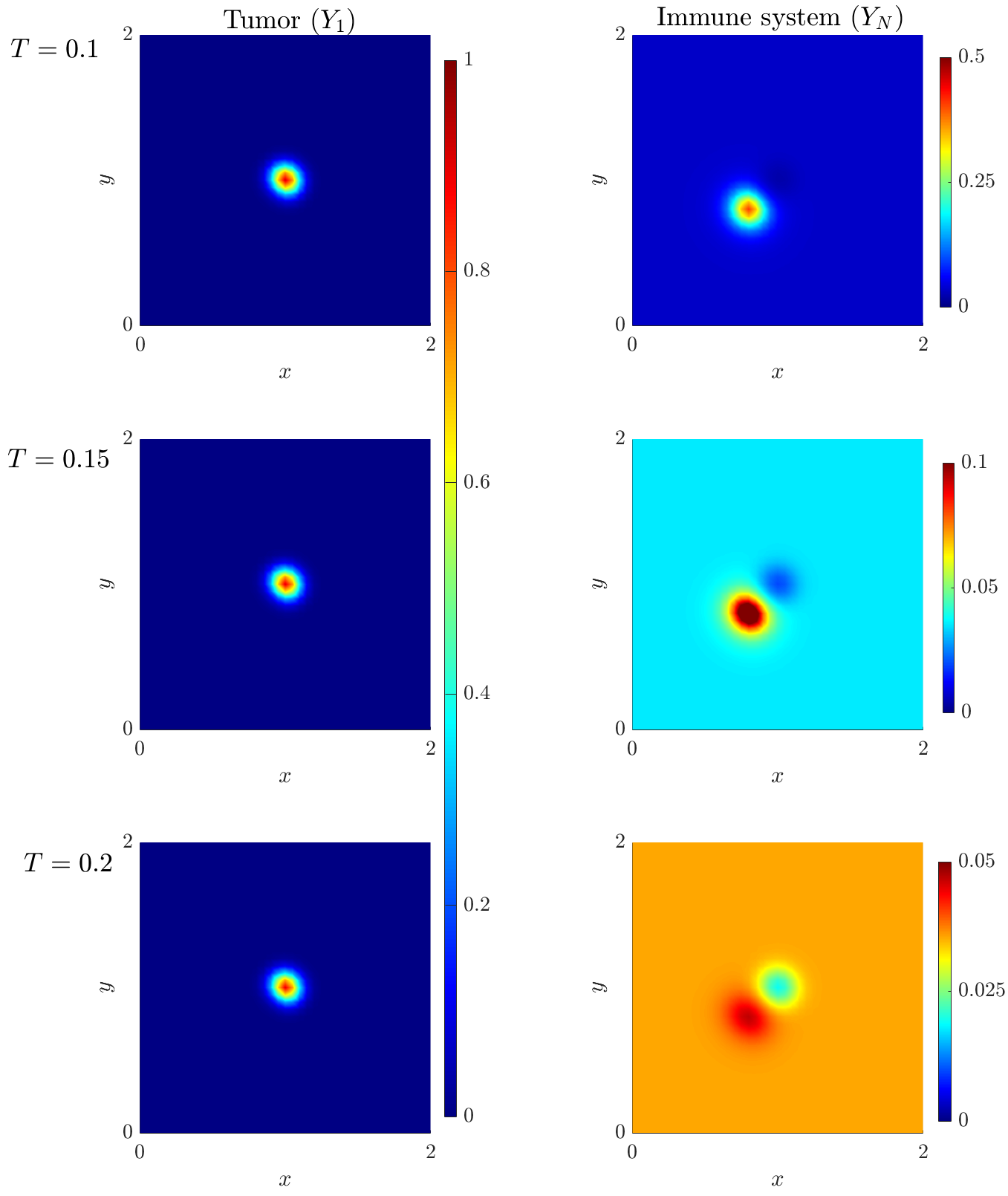}
    \caption{{\bf Test 2: evolution of system \eqref{Adim_Nonlin}.} Numerical simulation of model \eqref{Adim_Cross} with parameter values $A=G=50$, $B=D=1$, $C=7$, $E=F=35$, and $H=1.5$, and the diffusion coefficients $\mathcal{D}_1=10^{-6}$, $\mathcal{D}_2=0.01$, and $\mathcal{D}_3=0.0002$. The columns represent the tumor population ($Y_1$) and the total immune system ($Y_N$), respectively. The rows correspond to three different time steps: $T=0.1$, $0.15$, and $0.2$, respectively.}
    \label{2D_Cross2}
\end{figure}

\paragraph{{\bf Test 2: Proliferative scenario.}} We consider the case of proliferative dynamics for the immune populations. The adimensionalized version of system \eqref{S2_Lin} is given by: 
\begin{equation}\label{Adim_Lin2}
\begin{sistem}
      \dfrac{\partial \,Y_1}{\partial t}(t,\x)-\mathcal{D}_1 \Delta_\x \,Y_1(t,\x)=-A\,Y_1(t)\,Y_2(t,\x)+B\,Y_1(t,\x)\,Y_3(t,\x)+\,Y_1(t,\x)\,,\\[0.5cm]
     \dfrac{\partial \,Y_2}{\partial t}(t,\x)-\mathcal{D}_2 \Delta_\x \,Y_2(t,\x)=-A\,Y_1(t,\x)\,Y_2(t,\x)+D\,Y_3(t,\x)-G\,Y_2(t,\x)+H\,,\\[0.5cm]
      \dfrac{\partial \,Y_3}{\partial t}(t,\x)-\mathcal{D}_3 \Delta_\x \,Y_3(t,\x)=-C\,Y_3(t,\x)+E\,Y_2(t,\x)\,Y_3(t,\x)+F\,Y_2^2(t,\x)\,.
     \end{sistem}
\end{equation}
The adimensionalized version of system \eqref{NonLin_macro} is expressed as:

\begin{equation}\label{Adim_Nonlin}
 \begin{sistem}
     \dfrac{\partial \,Y_1}{\partial t}(t,\x)-\mathcal{D}_1 \Delta_\x \,Y_1(t,\x)=\dfrac{\,Y_1(t,\x)\,Y_N(t,\x)}{1+P\,Y_N(t,\x)}\Big[B\,P\,Y_N(t,\x)-A\Big]+\,Y_1(t,\x)\,,\\[0.7cm]
      \dfrac{\partial \,Y_N}{\partial t}(t,\x)-\nabla_\x\nabla_\x:\left[\dfrac{\mathcal{D}_2+\mathcal{D}_3 P\,Y_N(t,\x)}{1+P\,Y_N(t,\x)}\,Y_N(t,\x)\right]=\dfrac{\,Y_N(t,\x)}{1+P\,Y_N(t,\x)}\Big[-A\,Y_1(t,\x)+D\,P\,Y_N(t,\x)-G\Big]+H\,.
      \end{sistem}
 \end{equation}
Here, the parameters $D$, $E$, $F$, $G$ and $H$ are defined in Section \ref{OmoSisAnal_2}, and $\mathcal{D}_i:=\mathbb{D}_i\,(\delta_{14}\,n_4)^{-1}$ for $i=1,2,3$. For both systems, we use the same initial Gaussian-like aggregate of tumor cells defined by \eqref{InCon_Tum}. We assume that the immune system is entirely in an active state to promote immune cell proliferation. In system \eqref{Adim_Lin2}, the initial condition for the active immune cells is a Gaussian-like aggregate centered at  $(x_{0,\,Y_3}, y_{0,\,Y_3})=(0.8,0.8)$ with variance $\sigma^2_{\,Y_2}=0.2^{2}$:

$$
\,Y_{2,0}=\exp\left({-\dfrac{(x-x_{0,\,Y_2})^2+(y-y_{0,\,Y_2})^2}{\sigma_{\,Y_2}^{2}}}\right)\,,\qquad \,Y_{3,0}=0\,.
$$
For system \eqref{Adim_Nonlin}, we retain the same initial condition for the tumor cells and set the initial condition for the entire immune cell population as follows:

$$
\,Y_{N,0}=\exp\left({-\dfrac{(x-x_{0,\,Y_2})^2+(y-y_{0,\,Y_2})^2}{\sigma_{\,Y_2}^2}}\,\right).
$$
The results of the numerical simulations for models \eqref{Adim_Lin2} and \eqref{Adim_Nonlin} are presented in Figures \ref{2D_Lin2} and \ref{2D_Cross2}, respectively.

\begin{figure}[h!]
    \centering
    \includegraphics[width=.45\textwidth]{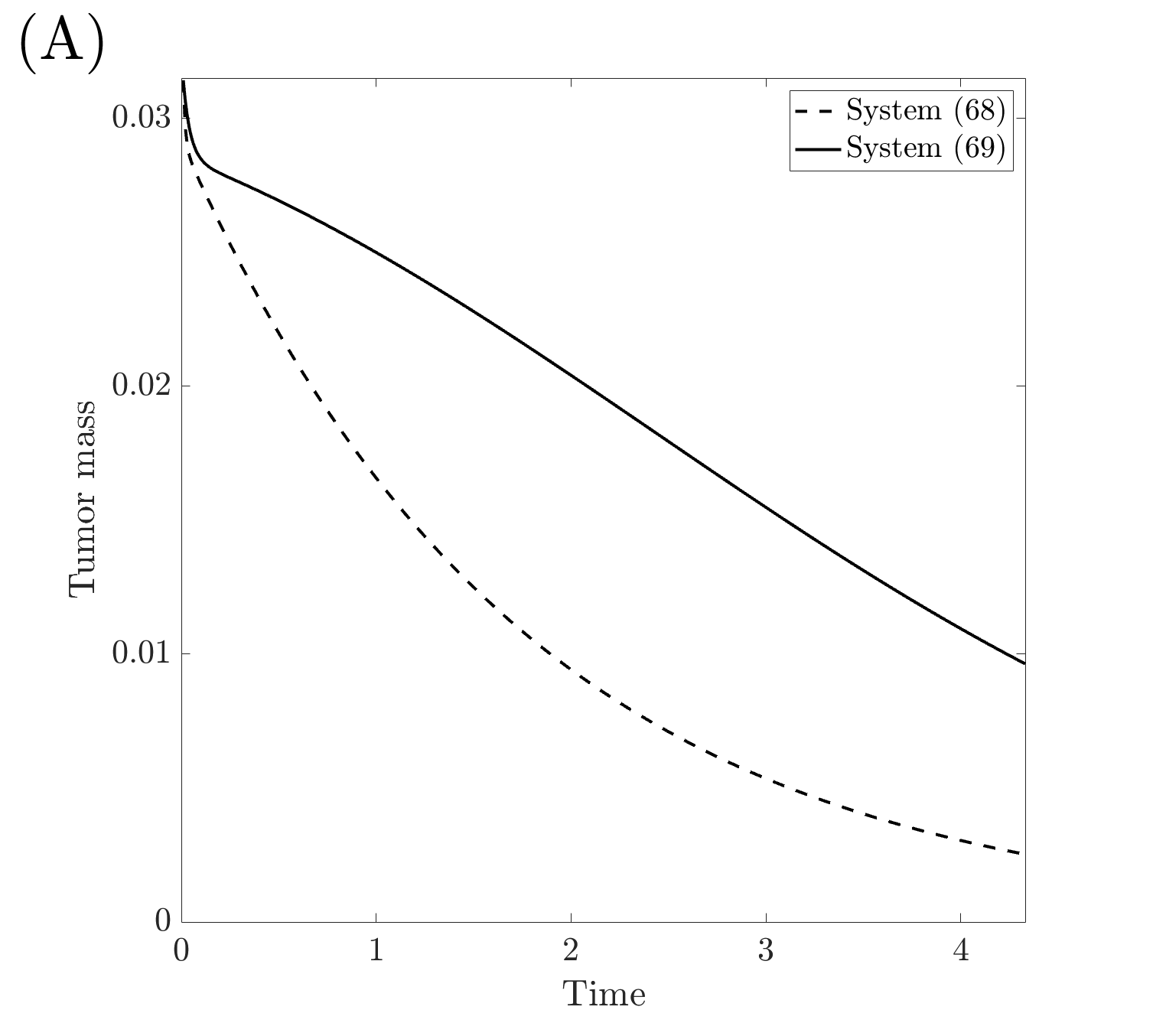}
    \includegraphics[width=.45\textwidth]{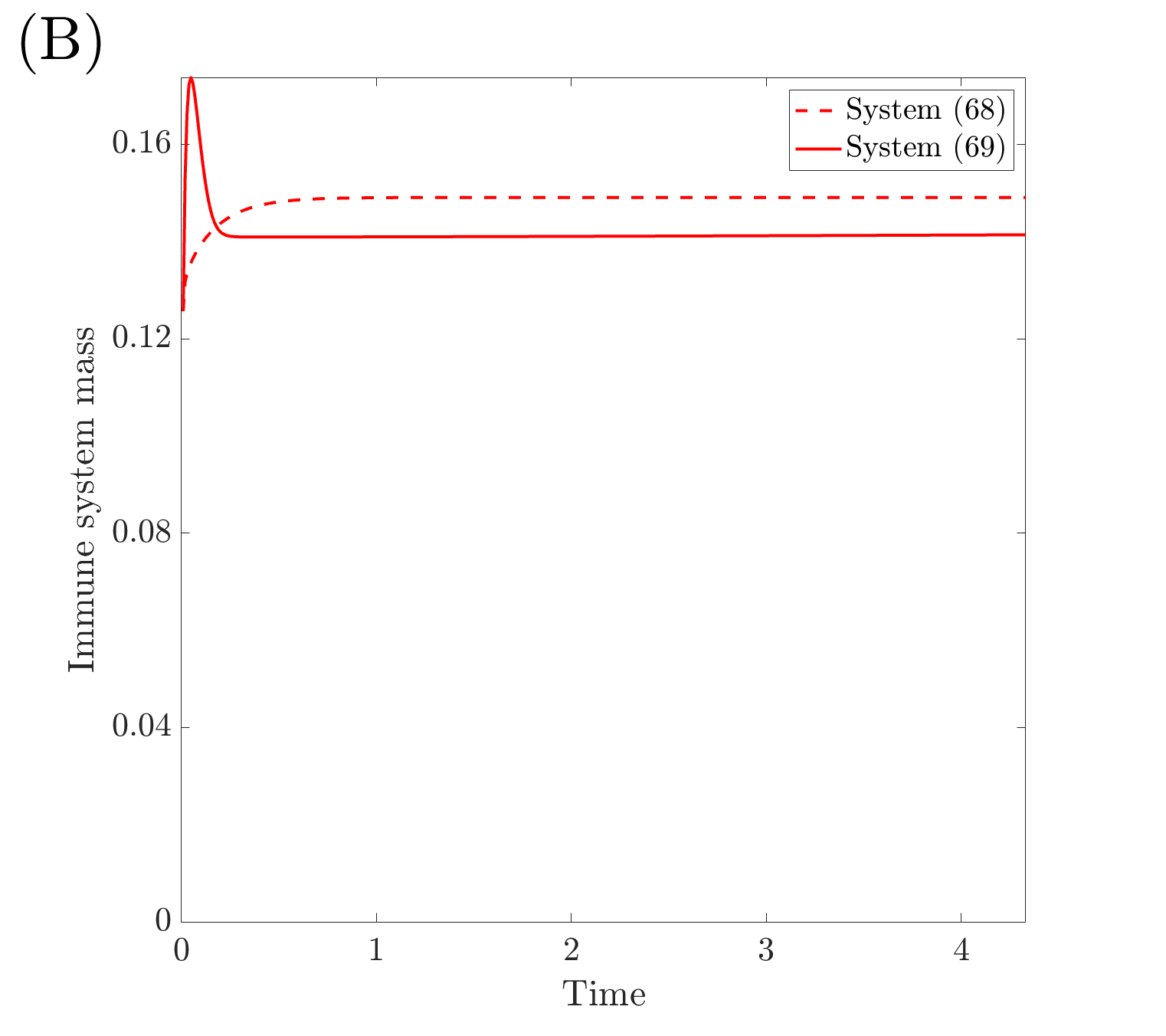}
    \caption{{\bf Test 2: temporal evolution of tumor and immune system masses.} Panel A: comparison of the total tumor mass over time for system \eqref{Adim_Lin2} (black dashed line) and system \eqref{Adim_Nonlin} (black solid line). Panel B: temporal evolution of the total immune system mass over time for system \eqref{Adim_Lin2} (red dashed line) and system \eqref{Adim_Nonlin} (red solid line). In both cases, the parameter values are set to $A=G=50$, $B=D=1$, $C=7$, $E=F=35$, and $H=1.5$, while the diffusion coefficients are set to $\mathcal{D}_1=10^{-6}$, $\mathcal{D}_2=0.05$, and $\mathcal{D}_3=0.0002$.}
    \label{Masses_prol}
\end{figure}

\noindent As observed in the previous case, both systems \eqref{Adim_Lin2} and \eqref{Adim_Nonlin} exhibit a qualitative decrease in tumor mass due to immune cell activity. However, unlike the earlier numerical tests, the dynamics here are considerably slower because the proliferation of the passive immune system supports tumor cell growth. When comparing the dynamics of the linear and nonlinear self-diffusion models, we find that the self-diffusion term in system \eqref{Adim_Nonlin} enhances the spreading of immune cells within the domain. Specifically, the equation for the total immune system population $Y_N$ in system \eqref{Adim_Nonlin} can be expressed as:

\begin{equation*}
\begin{split}
       &\dfrac{\partial \,Y_N}{\partial t}(t,\x)-\nabla_\x\cdot \left[\,Y_N(t,\x)\nabla_\x\left(\dfrac{\mathcal{D}_2+\mathcal{D}_3 P\,Y_N(t,\x)}{1+P\,Y_N(t,\x)}\right)\right]-\nabla_\x\cdot\left[\dfrac{\mathcal{D}_2+\mathcal{D}_3 P\,Y_N(t,\x)}{1+P\,Y_N(t,\x)}\,\nabla_\x \,Y_N(t,\x)\right]\\[0.2cm]
      &=\dfrac{\,Y_N(t,\x)}{1+P\,Y_N(t,\x)}\Big[-A\,Y_1(t,\x)+D\,P\,Y_N(t,\x)-G\Big]+H\,.
\end{split}
\end{equation*}
This increased diffusivity results in an even slower process for tumor cell elimination. This behavior is evident when comparing Figure \ref{Masses_prol}-A with Figure \ref{Masses_cons}-A. Furthermore, the dynamics also affect the immune population in system \eqref{Adim_Nonlin}. This population takes significantly longer to reach its equilibrium value compared to the immune population in \eqref{Adim_Lin2}, which achieves its relaxation state more rapidly (see Figure \ref{Masses_prol}-B).

\section{Concluding remarks and perspectives}\label{SecConc}
In this work, we have developed a comprehensive mathematical framework to elucidate the diverse interaction mechanisms between tumor cells and the immune system, as these interactions play a crucial role in the dynamics of various tumor types \cite{korell2022understanding}. For instance, in the case of brain tumors, understanding the spatial dynamics between glioma cells and T cells is essential for optimizing immunotherapy treatments \cite{maggs2021car}. Our main result is the derivation of macroscopic systems from kinetic models, incorporating different types of diffusive operators that reflect the distinct migratory behaviors of cells. Employing a kinetic approach rooted in the classical methods of the kinetic theory of active particles, we defined two distinct spatially distributed models of tumor-immune dynamics. In these models, each cell is characterized by its position $\x$, microscopic velocity $\bv$, and internal state $u$. Specifically, the first model, presented in \eqref{case1_Boltz}, describes a conservative scenario for the immune system, where immune cells undergo activation (via interleukin boosting) or deactivation processes and spatial reorientation. In this context, tumor cells may proliferate or die as a result of their interactions with the immune cells. In contrast, the second model, outlined in \eqref{case2_Boltz}, incorporates immune system proliferation and apoptosis, providing a more dynamic representation of immune responses within the tumor microenvironment. By considering different time scales for the processes involved, we were able to derive macroscopic systems exhibiting either linear or nonlinear diffusion through a formal hydrodynamic limit, as shown in Section \ref{deri_macroSys}. From the conservative setting described in \eqref{case1_Boltz}, we derived two macroscopic frameworks: one featuring linear diffusion for all species (system \eqref{S1_Lin}) and the other showcasing nonlinear cross-diffusion for the immune populations (system \eqref{CrosDif_macro}). Conversely, from the proliferative setting in \eqref{case2_Boltz}, we established two additional macroscopic frameworks: one with linear diffusion for all species (system \eqref{S2_Lin}) and another with nonlinear self-diffusion for the immune populations (system \eqref{NonLin_macro}).

The derived macroscopic systems were qualitatively analyzed in spatially homogeneous scenarios using classical tools from the theory of dynamical systems. Our primary objective was to gain insights into the existence and stability of potential equilibrium configurations. Through this analysis, we discovered that, in both conservative and proliferative cases, the different scaling used in deriving the macroscopic models from the kinetic settings does not significantly impact the asymptotic behavior of the homogeneous components. Instead, these characteristics are primarily determined by the kinetic structure of the models themselves. Both the qualitative simulations of the homogeneous settings in Section \ref{Homog_section} and those of the spatially distributed configurations in Section \ref{Macro_sim} demonstrate that the varying scaling applied during the macroscopic derivation primarily influences the transient phase and, specifically, the time required to reach equilibrium configurations. Notably, the slower dynamics associated with the derivation of systems featuring nonlinear diffusion result in extended stabilization periods for these systems. The numerical simulations of the spatially distributed settings further underscored - even in simplified scenarios where heterogeneity is not considered - the substantial role of spatial dynamics in shaping system evolution. Specifically, we observed more confined spreading when the immune system features cross-diffusion, and broader diffusion when immune cell motion is governed by self-diffusion. These results emphasize the importance of deepening our understanding of the biological mechanisms that drive immune cell migration, an topic that still presents significant knowledge gaps; for instance, T cell migration in response to brain tumor progression is not sufficiently understood yet. Gaining deeper insight into this area could be particularly relevant for optimizing treatment strategies aimed at controlling tumor growth. Additionally, our analysis reveals that, while the conservative case yields relatively simple dynamics, characterized by two possible equilibrium configurations with well-defined stability conditions, the introduction of proliferation leads to significantly more complex behavior and a wider range of possible outcomes. Notably, we observe the emergence of a stable limit cycle around a coexistence equilibrium configuration, indicating sustained oscillatory dynamics. This phenomenon is particularly significant, as it reflects oscillations between tumor cells and therapeutic agents, a pattern that has also been observed in similar modeling contexts \cite{conte2025car, sardar2021exploring}.  Such dynamics are consistent with clinical observations of tumor dormancy and relapse, in which cancer cells remain viable but non-proliferative, eventually leading to disease recurrence following a period of remission \cite{tufail2025tumor}. Importantly, across all considered scenarios, we identified a specific set of parameters that ensures the stability of a disease-free equilibrium configuration.

The diverse range of models and analyses proposed in this work opens several avenues for further investigation, aimed at deepening our understanding of tumor-immune interactions and enhancing therapeutic strategies. Firstly, we plan to extend the derived approaches by introducing environmental heterogeneity into the spatially distributed macroscopic settings. This will involve incorporating space-dependent diffusion coefficients to assess their influence on spatial dynamics across different scenarios. Additionally, we will include heterogeneity in the static population ($n_4(\x)$ and $n_5(\x)$), which are currently assumed to be spatially homogeneous. This inclusion aims to develop more realistic models that account for the limited resources available to tumor cells and the effects of therapy administration, such as localized injections. By introducing spatial variability, he complexities of the tumor microenvironment and its interactions with the immune system can be better captured. Another promising direction is the formulation and analysis of an optimal control problem related to the therapeutic terms in our models. This will help identify the optimal spatial and temporal conditions for interleukin administration, potentially improving treatment efficacy. Given the generality of our framework, it can also be adapted to study other pathologies involving immune responses, such as autoimmune diseases or chronic inflammatory conditions, thus broadening the applicability of the developed models. Furthermore, we recognize that some of the assumptions used to model the biological processes (such as cell proliferation or cell turning), despite being consistent with existing literature \cite{conte2018qualitative, bellomo2008mathematical}, may not fully reflect their biological complexity; thus, we plan to investigate the outcomes that can be obtained under more realistic hypotheses.

In summary, the proposed mathematical kinetic framework serves as a novel and valuable tool for studying various spatial interactions between tumor and immune cells. Our findings provide insightful interpretations of the asymptotic dynamics that emerge within these systems, paving the way for both theoretical advancements and practical applications in the field of tumor-immune interactions.

\subsection*{Acknowledgements} 
The authors would like to thank Dr. Giorgio Martalò for helpful discussions. 
The research of both authors has been carried out under the auspices of GNFM (National Group of Mathematical-Physics) of INDAM (National Institute of Advanced Mathematics). MC was supported by the National Group of Mathematical Physics (GNFM-INdAM) through the INdAM–GNFM Project (CUP E53C22001930001) {\it ‘From kinetic to macroscopic models for tumor-immune system competition’} and by the European Union and the Italian Ministry of University and Research through the PNRR project Young Researchers 2024—SOE {\it ‘Integrated Mathematical Approach to Tumor Interface Dynamics’} (CUP: E13C24002380006). MC has been partially supported by the State Research Agency of the Spanish Ministry of Science and FEDER-EU, project PID2022-137228OB-I00 (MICIU/AEI /10.13039/501100011033); by Modeling Nature Research Unit, Grant QUAL21-011 funded by Consejería de Universidad, Investigaci\'on e Innovaci\'on (Junta de Andalucía). The work has been performed in the frame of the project PRIN 2022 PNRR {\it ‘Mathematical Modelling for a Sustainable Circular Economy in Ecosystems’} (project code P2022PSMT7, CUP D53D23018960001) funded by the European Union - NextGenerationEU and by MUR-Italian Ministry of Universities and Research. RT is a post-doc fellow supported by the National Institute of Advanced Mathematics (INdAM), Italy. RT was supported by the Portuguese national funds (OE), through FCT/MCTES Projects UIDB/00013/2020, UIDP/00013/2020, PTDC/03091/2022 ({\it ‘Mathematical Modelling of Multi-scale Control Systems: applications to human diseases (CoSysM3)’}). RT also thanks the support of the University of Parma through the action Bando di Ateneo per la ricerca 2022 co-funded by MUR-Italian Ministry of Universities and Research - D.M. 737/2021 - PNR - PNRR - NextGenerationEU (project: {\it ‘Collective and self-organised dynamics: kinetic and network approaches’}).

\subsection*{Author contribution statement}
All authors contributed equally to the article and approved the submitted version.

\subsection*{Data availability statement}
Data sharing is not applicable to this article as no data sets were generated or analysed during the current study.

\subsection*{Conflict of interest}
The authors declare that the research was conducted in the absence of any commercial or financial relationships that could be construed as a potential conflict of interest.

\newcommand{\noopsort}[1]{}
\nocite{*}
\bibliographystyle{abbrv}
\bibliography{Bibliography}

\end{document}